\documentclass[10pt,aps,prc,twocolumn,showpacs,showkeys,amsmath,floatfix,superscriptaddress]
{revtex4-1}
\usepackage{color,graphicx}
\usepackage{mathptmx}                
\usepackage{dcolumn}                 
\usepackage{bm}                      
\usepackage{ulem,soul}

\begin{document}


\title{Microscopic evaluation of the hypernuclear chart with $\Lambda$-hyperons}

\author{E. Khan}
\affiliation{Institut de Physique Nucl\'eaire, 
Universit\'e Paris-Sud, IN2P3-CNRS, F-91406 Orsay c\'edex, France}
\author{J. Margueron}
\affiliation{Institut de Physique Nucl\'eaire de Lyon, 
Universit\'e Claude Bernard Lyon 1, \\IN2P3-CNRS, F-69622 Villeurbanne c\'edex, France}
\author{F. Gulminelli}
\affiliation{CNRS/ENSICAEN/LPC/Universit\'e de Caen Basse Normandy, 
UMR6534, 
F-14050 Caen c\'edex, France}
\author{Ad. R. Raduta}
\affiliation{IFIN-HH, Bucharest-Magurele, POB-MG6, Romania}

\begin{abstract}
\begin{description}
\item[Background] 
A large number of hypernuclei, where  a considerable fraction of nucleons is replaced by strange baryons, and even
pure hyperonic species are expected to be bound. Though, the hypernuclear landscape remains largely unknown because
of scarce constraints on the $NY$ and $YY$ interactions.
\item[Purpose] 
We want to estimate the number of potentially bound hypernuclei. 
In order to  evaluate realistic error bars within the theoretical uncertainties associated to the spherical mean-field approach,
and the present information from already synthetized hypernuclei 
on the $N-Y$ and $Y-Y$ channels,
we limit ourselves to purely $\Lambda$-hypernuclei, to magic numbers of $\Lambda$'s 
(for Z $\leq$ 120 and $\Lambda \leq$70), and to even-even-even systems.
\item[Method] We consider a density functional approach adjusted to microscopic Bruckner-Hartree-Fock calculations,
where the $\Lambda\Lambda$ term is corrected in a phenomenological
way, to reproduce present experimental constraints.  Different models
which strongly deviate at large densities, but giving the same bond
energy, are generated in order to take into account the uncertainties
related to the high density equation of state.
\item[Results] 
The number of bound even-even-even $\Lambda$-hypernuclei  is estimated to 491680 $\pm$ 34400. 
This relatively low uncertainty is due to the fact that the well constrained low density and highly unconstrained high density behavior
of the energy functional turn out to be largely decoupled.
Results in $\Lambda$-hypernuclei appear to be almost independent of the choice for the high-density part of the $\Lambda\Lambda$ interaction.
The location of the $\Lambda$-hyperdriplines is also evaluated. Significant deviations from
Iron-Nickel elements can be found for $\Lambda$-hypernuclei with the largest
binding energy per baryon. Proton, neutron and $\Lambda$-hyperon magicity
evolution and triple magic $\Lambda$-hypernuclei are studied. Possible bubbles
and haloes effect in $\Lambda$-hypernuclei are also discussed.
\item[Conclusions] 
The present results provide a first microscopic evaluation of the
$\Lambda$-hypernuclear landscape. They shall benefit from the more and more accurate
design of the $\Lambda$-based functionals. 
The measurements of
$\Lambda$ and multi-$\Lambda$ hypernuclei together with additional constraints of 
the $\Xi\Lambda$ and $\Xi\Xi$
interaction is mandatory to improve such critical information.
\end{description}

\end{abstract}

\date{\today}


\maketitle

\section{Introduction}

The study of hypernuclei benefited from a great scientific
interest since the 
'60s \cite{Bodmer1964,Dalitz1976}.
Various hyperon-nucleus interaction
\cite{Millener1988}, 
Skyrme Hartree-Fock
\cite{Rayet1976, Auerbach1977,lan98,meng06,zho07,hon08,gul12,min12},
relativistic mean-field
\cite{Bouyssy1982,Rufa1987,Mares1989,ruf90,Schaffner1992,Mares1993,Schaffner1994,Sedrakian2014},
generalized liquid drop
\cite{Balberg_masses_1994,gal_GBW1993,Samanta2006,Samanta2008, Samanta2010}
and, G-matrix~\cite{lan97,cug00,vid01} models have
been considered in order to underpin the hyperon-nucleon interaction from hypernuclei,
to test the existence of bound hypernuclei, the stability of nucleonic cores
against hyperon addition, or the occurrence of exotic pure strange systems or
halo structures.
With the advent of new dedicated experimental
programs such as J-Parc in Japan or PANDA at FAIR, the study of the
hypernuclei structure enjoys a revived interest
\cite{min13,ikr14,san14}. 

Present facilities can only produce single and double $\Lambda$ 
hypernuclei in a limited domain of mass. However, a general
understanding of the specificity of hypernuclear structure with
respect to nuclear structure requires the evaluation of the global
hypernuclear chart, with strangeness as the third dimension
\cite{gre01,Samanta2006}. 
Some
recent works \cite{ikr14,san14} address this problem within
phenomenological mean field models. However, the uncertainties on the
hypernuclear chart associated to the choice of the functional in the
strangeness sector are difficult to evaluate, since present
experimental data on hypernuclei are scarce, and therefore a large
arbitrariness is associated to the modeling of the hyperonic
energy functional. To minimize such uncertainties, it is important to
use as much as possible microscopically founded energy functionals
from Brueckner or Dirac-Brueckner calculations
as well as the few available experimental data. 
The present work is
an attempt towards that direction.

The regular nuclear landscape, including an estimation of the
uncertainties on its limits defined by the drip-lines, has been only
recently microscopically studied \cite{erl12}, owing to the growth of
calculation capacities. This is also due to the recent use of evaluation
methods \cite{rei13,che14} for the uncertainties generated by the only
input of such microscopic calculations: the nucleon-nucleon (NN)
energy functional. The resulting uncertainty on the number of bound
nuclei is typically of the order of 7\% \cite{erl12}.

The present work aims to generalize such a study by designing the
limits of the $\Lambda$-hypernuclear landscape, and evaluating the uncertainties
associated with the limited empirical information on the hyperonic functionals.
For this purpose microscopic
calculations of nuclei and $\Lambda$-hypernuclei are performed using the energy
density functional approach. We use a Skyrme functional optimized on nuclear physics data
for the NN channel, while $\Lambda$N and $\Lambda$$\Lambda$ functionals
fitted on microscopic Brueckner-Hartree-Fock (BHF) calculations are used, 
with modifications in order to take into account
$\Lambda$-hypernuclei data constraints.

When dealing with multi-strange system, a very important issue is
the possible presence of hyperons other than $\Lambda$. 
In particular,  it was recognized since the early 
nineties~\cite{Schaffner1992,Schaffner1994,Schaffner2000} 
that the $\Lambda + \Lambda \rightarrow \Xi+N$ decaying channel could play 
a major role in multi-strange systems, because of the attractive character of 
the $\Xi$ potential in nuclear matter, which could lead to the appearance of 
$\Xi$ hyperons already for a strangeness number $|S|>8$.
The present most accepted value for 
the $\Xi$  potential in symmetric nuclear matter $V_\Xi=-14$ MeV~\cite{Khaustov2000} 
is much less attractive 
than the value proposed in those earlier works~\cite{Schaffner1994},
which pushes the $\Lambda$ number threshold for $\Xi$ contribution farther in strangeness. 
Moreover, it is clear that the contribution of $\Xi$ will crucially depend
on the $\Xi$-$\Xi$ and $\Xi$-$\Lambda$ interactions, which are presently completely unconstrained. 
The inclusion or removal of the Fock term in relativistic approaches seams to have, as well, an effect on the
appearance the various kinds of hyperons in uniform matter~\cite{Massot2012}.
Because of these uncertainties, we have chosen to limit ourselves to $\Lambda$-
hypernuclei in this work, since $\Lambda$'s are the only hyperons to be 
(relatively) constrained by experimental data.

Section II details the energy density functional used, focusing on
the way to design the $\Lambda$$\Lambda$ component.  
Section III is devoted to the determination of the
corresponding parameters of the $\Lambda$$\Lambda$ functional. The
Hartree-Fock calculations for $\Lambda$-hypernuclei are performed in Section IV.
Since in the hyperonic channel the spin-orbit interaction is expected 
to be very weak~\cite{fin09},
the $\Lambda$-hypernuclear charts, location of the $\Lambda$-hyperdriplines and
the estimation of the number of bound
even-even-even $\Lambda$-hypernuclei are evaluated
for hypernuclear number $\Lambda$=0, 2, 8, 20, 40 and 70.
In Section V, the gross properties of
$\Lambda$-hypernuclear structure are analyzed, namely the evolution of the energy
per baryon as a function of the $\Lambda$N and $\Lambda$$\Lambda$
functionals. A study of magicity evolution in $\Lambda$-hypernuclei, as well as
of possible bubbles and haloes effects is also undertaken. Finally,
Appendix A details the link between the bond energy and the
$\Lambda$$\Lambda$ functional whereas Appendix B provides an update of
the strangeness analog resonances in multihyperons $\Lambda$-hypernuclei.

\section{Density functional theory for $\Lambda$-hyper-nuclear matter and $\Lambda$-hyper-nuclei}

We consider a non-relativistic system composed of interacting nucleons $N$ and lambdas $\Lambda$.
The total Hamiltonian reads,
\begin{eqnarray}
\hat{H}=\hat T_N +\hat T_\Lambda + \hat H_{NN} + \hat H_{\Lambda\Lambda} + \hat H_{N\Lambda}, 
\label{ham} 
\end{eqnarray}
where $\hat T_A$ is the kinetic energy operator and $\hat H_{AB}$ the interaction operator acting between 
$A$ and $B$ (=$N$ and $\Lambda$).
We work in the mean-field approximation where the ground state
of the system is given by the tensor product, $\vert\Phi_N\rangle\otimes\vert\Phi_\Lambda\rangle$, where 
$\vert\Phi_N\rangle=\Pi_i a^+_i\vert-\rangle$ is a Slater determinant of nucleon states and 
$\vert\Phi_\Lambda\rangle=\Pi_\lambda a^+_\lambda\vert-\rangle$ is a Slater determinant of lambda states. 
The total Hamiltonian~(\ref{ham}) can be expressed in terms of the nucleons ($i$) and lambda ($\lambda$) states as,
\begin{eqnarray}
\hat{H}=\sum_i \hat t_i + \sum_\lambda \hat t_\lambda + \frac{1}{2} \sum_{i,j} \hat{v}_{ij}^{NN} + \frac{1}{2} \sum_{\lambda,\mu} \hat{v}_{\lambda\mu}^{\Lambda\Lambda}+ \sum_{\lambda ,i}  \hat{v}_{i\lambda}^{N\Lambda} 
\label{ham2} .
\end{eqnarray}

In the following, we will consider the density functional theory
which allows 
relating in a direct way the BHF predictions for uniform matter to
the properties of hyper-nuclei.

\subsection{Energy-density functional deduced from BHF}\label{sec:func}

In the present study of $\Lambda$-hypernuclei and nuclear matter we use a density functional which has been determined
directly from BHF theory including nucleons and $\Lambda$-hyperons~\cite{cug00,vid01}.
The total energy density $\epsilon(\rho_N,\rho_\Lambda)$ is related to the energy per particle calculated within
the BHF framework, $e_{BHF}$, as $\epsilon(\rho_N,\rho_\Lambda)=(\rho_N+\rho_\Lambda) e_{BHF}(\rho_N,\rho_\Lambda)$
and is decomposed in different terms,
\begin{eqnarray}
\epsilon(\rho_N,\rho_\Lambda) &=& \frac{\hbar^2}{2m_N}\tau_N+ \frac{\hbar^2}{2m_\Lambda}\tau_\Lambda+\epsilon_{NN}(\rho_N) \nonumber \\
&&\hspace{1cm}+\epsilon_{N\Lambda}(\rho_N,\rho_\Lambda)
+\epsilon_{\Lambda\Lambda}(\rho_\Lambda) ,
\label{functional}
\end{eqnarray}
where, in infinite nuclear matter, the kinetic energy densities $\tau_N$ and $\tau_\Lambda$ are simple 
functions of the matter density: $\tau_i= \frac 35  ( 6\pi^2/g_i
)^{2/3} \rho_i^{5/3}$, with $g_i=4(2)$
for $i=N(\Lambda)$.

In the nucleon sector, we use the SLy5 parametrization of the 
phenomenological Skyrme functional including non-local and spin-orbit
terms, since it can correctly reproduce the properties of stable and
exotic nuclei~\cite{cha98}. In the case of the strangeness sector, the
spin-orbit interaction is known to be small~\cite{fin09}, and is
therefore neglected. The local density dependence of the
$N\Lambda$ component of the energy density,
$\epsilon_{N\Lambda}(\rho_N,\rho_\Lambda)$, is solely adjusted to the
BHF predictions. To pin down the $N\Lambda$ coupling, the
following energy density is defined: $(\rho_N+\rho_\Lambda)
e_{BHF}(\rho_N,\rho_\Lambda)-\rho_N e_{BHF}(\rho_N,0)-\rho_\Lambda
e_{BHF}(0,\rho_\Lambda)$,  where $ e_{BHF}(\rho_N,\rho_\Lambda)$
is the BHF energy per baryon in an infinite hyper-nuclear matter
calculation. It is parameterized in terms of the nucleon and hyperon densities
as~\cite{cug00,vid01},
\begin{eqnarray}
\epsilon_{N\Lambda}(\rho_N,\rho_\Lambda) &=& -f_1(\rho_N) \rho_N\rho_\Lambda + f_2(\rho_N)\rho_N\rho_\Lambda^{5/3},
\label{eq:enl}
\end{eqnarray}
where the first term physically corresponds to the attractive $N\Lambda$ interaction,
corrected by the presence of the medium given by the function
$f_1$, and the second term is induced by the repulsive momentum dependent term
of the $\Lambda$ potential (considering the low-momentum quadratic
approximation), also corrected by the medium through the function
$f_2$. In the presence of the attractive $\Lambda\Lambda$ interaction,
the term $\epsilon_{\Lambda\Lambda}$ is solely determined by the
hyperon density as~\cite{vid01},
\begin{equation}
\epsilon_{\Lambda\Lambda}(\rho_\Lambda)=-f_3(\rho_\Lambda)\rho_\Lambda^2 .
\label{e_LL}
\end{equation}

The functions $f_i$ are given by the polynomial forms,
\begin{eqnarray}
f_1(\rho_N) &=& \alpha_1-\alpha_2\rho_N+\alpha_3\rho_N^2 , \label{eq:fi1}\\
f_2(\rho_N) &=& \alpha_4-\alpha_5\rho_N+\alpha_6\rho_N^2 , \\
f_3(\rho_\Lambda) &=&\alpha_7-\alpha_8\rho_\Lambda+\alpha_9\rho_\Lambda^2.
\label{eq:fi}
\end{eqnarray}

\begin{table*}[t]
\caption{Parameters of the f$_i$ functions, see Eqs.~(\ref{eq:fi1})-(\ref{eq:fi}), for the functionals DF-NSC89, DF-NSC97a, DF-NSC97f.}
\begin{ruledtabular}
\begin{tabular}{l|ccccccccccc}
Force & $\alpha_1$ & $\alpha_2$ & $\alpha_3$ & $\alpha_4$ & $\alpha_5$ & $\alpha_6$ & $\alpha_7$ & $\alpha_8$ & $\alpha_9$ \\
\hline
DF-NSC89~\cite{cug00,vid01} & 327 & 1159 & 1163 & 335 & 1102 & 1660 & 0 & 0 & 0 \\
DF-NSC97a~\cite{vid01} & 423 & 1899 & 3795 & 577 & 4017 & 11061 & 38 & 186 & 22 \\
DF-NSC97f~\cite{vid01} & 384 & 1473 & 1933 & 635 & 1829 & 4100 & 50 & 545 & 981 \\
\end{tabular}
\end{ruledtabular}
\label{table:SZ1}
\end{table*}

In principle, the functions $f_1$ and $f_2$ depend on the densities
$\rho_N$ and $\rho_\Lambda$ associated to conserved charges in the medium.
However, since nucleons are the dominant species, even in the presence of lambdas,
the dependence on $\rho_\Lambda$ is neglected in these functions.
In the case of $f_3$, it trivially depends on $\rho_\Lambda$ only, because
it impacts the
part of the functional referring to pure lambda matter.

Different $N\Lambda$ potentials have been parameterized which equally
well fit the scarce $N\Lambda$ phase shifts, see for instance
discussion and references in Ref.~\cite{Schulze2013}. The present
study is based on three of them for which a density functional has been
derived, based on BHF predictions~\cite{cug00,vid01}, namely DF-NSC89,
DF-NSC97a and DF-NSC97f. The functional DF-NSC89 is based on the
Nijmegen soft-core potential NSC89~\cite{Maessen1989}, while the
functionals DF-NSC97a and DF-NSC97f are based on two of a series of six different
hyperon-nucleon potentials which 
equally well reproduce the measured scattering
lengths in the $\Lambda N$ and $\Sigma N$
channels, and correspond to different values 
of the magnetic vector  meson ratio $\alpha$ which 
cannot be constrained by the phase shift information ~\cite{Rijken1999,Stoks1999}.  
 
Although more recent functionals have been
parameterized~\cite{Schulze2013}, by adding for instance isospin
degree of freedom, the functionals considered in this study already
represent a good sample of the uncertainty generated by the lack of
empirical information in the strangeness sector. This affects 
microscopic approaches like BHF~\cite{cug00,vid01,zho07}, although to
a lesser extent than fully phenomenological mean field
models~\cite{ruf90,Schaffner1992,Schaffner1994,Mares1993,lan97,lan98,
meng06,hon08,gul12,min12,min13,san14,ikr14}.
Indeed the models DF-NSC97a and DF-NSC97f correspond to the two
extreme choices for the unconstrained $\alpha$
parameter~\cite{Rijken1999,Stoks1999}, leading to the softest and
stiffest equation of state, respectively. It should be noted that no
experimental information is available on $\Lambda$-$\Lambda$
scattering, meaning that these phenomenological bare interactions are
completely unconstrained in the $\Lambda$-$\Lambda$ channel. For this
reason, NSC89 does not contain any $\Lambda$-$\Lambda$ interaction.
The NSC97 models assume for this channel a simple SU(3) extension of
the original Nijmegen potential models to multiple strangeness $S=-2$.
The values for the parameters $\alpha_1-\alpha_9$ of the functions
$f_1-f_3$ are given in Table~\ref{table:SZ1} for the functionals
DF-NSC89, DF-NSC976a and DF-NSC97f.  

The single particle energies in uniform matter are deduced from the
energy functional~(\ref{functional}), as
$e_N^{unif}(k)=\frac{\hbar^2 k^2}{2m_N}+v_N^{unif}$ and
$e_\Lambda^{unif}(k)=\frac{\hbar^2 k^2}{2m_\Lambda}+v_\Lambda^{unif}$,
where
\begin{equation}
v_N^{unif}(\rho_N,\rho_\Lambda) = v_N^{sky}+\frac{\partial \epsilon_{N\Lambda}}{\partial \rho_N},
\end{equation}
$v_{N}^{Sky}$ being deduced from the Skyrme functional \cite{Bartel2008,ben03},
and
\begin{equation}
v_\Lambda^{unif}(\rho_N,\rho_\Lambda) = \frac{\partial \epsilon_{N\Lambda}}{\partial \rho_\Lambda}
+\frac{\partial \epsilon_{\Lambda\Lambda}}{\partial \rho_\Lambda}.
\label{eq:vlunif}
\end{equation}

In the following, the hyperon potential $v_{\Lambda}^{unif}$ is decomposed into
two terms,
$v_{\Lambda}^{unif}=v_{\Lambda}^{(N), unif}+v_{\Lambda}^{(\Lambda),
unif}$ using Eqs. (\ref{eq:enl}) and (\ref{e_LL}) ,

\begin{eqnarray}
v_{\Lambda}^{(N), unif}&=&\frac{\partial \epsilon_{N\Lambda}}{\partial \rho_\Lambda}
=-f_1(\rho_N)\rho_N+\frac 5 3 f_2(\rho_N) \rho_N \rho_\Lambda^{2/3} \label{eq:VLNunif} \\
v_{\Lambda}^{(\Lambda), unif}&=&\frac{\partial \epsilon_{\Lambda\Lambda}}{\partial \rho_\Lambda}
=-2\alpha_7\rho_\Lambda+3\alpha_8\rho_\Lambda^2-4\alpha_9\rho_\Lambda^3   .
\label{eq:VLLunif}
\end{eqnarray}

\begin{figure}[tb]
\begin{center}
\includegraphics[width=0.5\textwidth]{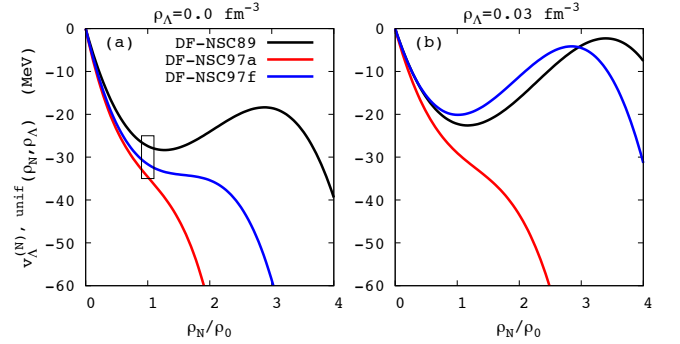}
\caption{(Color online) Potential $v_\Lambda^{(N), unif}(\rho_N,\rho_\Lambda)$ as a function of the nucleon density $\rho_N$ 
(in units of the saturation density $\rho_0$) for the functionals DF-NSC89, DF-NSC97a and
DF-NSC97f without $\Lambda$ (a) and with  $\rho_\Lambda=0.03$~fm$^{-3}$ (b).}
\label{fig:vln}
\end{center}
\end{figure}

The term $v_{\Lambda}^{(N), unif}$ stands for the contribution of the
nucleons to the hyperon potential, while the term
$v_{\Lambda}^{(\Lambda), unif}$ represents the direct contribution of
the hyperons to their own potential. The properties of the potential
$v_{\Lambda}^{(N), unif}$ are analyzed in Fig.~\ref{fig:vln} imposing
$\rho_n=\rho_p$.  On the left panel, the potential $v_{\Lambda}^{(N),
unif}$ is displayed without $\Lambda$ particles, while on the right
panel a small amount of $\Lambda$ is considered, corresponding to a
representative average $\Lambda$-density in single $\Lambda$
hypernuclei (see Sec.~\ref{sec:ldensitynuclei}). It is expected, from
experimental single $\Lambda$-hypernuclei data, that the potential
$v_{\Lambda}^{(N)}$ is about -30 MeV at saturation density
\cite{cug00}. This empirical condition is satisfied for the three
functionals DF-NSC89, DF-NSC97a and DF-NSC97f at saturation density,
as shown on the left panel of Fig.~\ref{fig:vln}.  For a fixed and
small amount of $\Lambda$, the potential $v_{\Lambda}^{(N), unif}$ is
attractive at large densities for all the functionals, due to the
$\alpha_3$ term in the function $f_1$.  
Already in the late eighties it was clear that  some
repulsion is necessary at high densities to explain hypernuclear data ~\cite{Millener1988}.
For a finite amount of
$\Lambda$, the $\alpha_6$ term in function $f_2$ gives a repulsive
contribution, and can compensate the attractive $\alpha_3$ term if 
$\rho_\Lambda>\left(\frac{3}{5}\frac{\alpha_3}{\alpha_6}\right)^{3/2}$.
For the functionals DF-NSC89, DF-NSC97a and DF-NSC97f, this occurs for
$\rho_\Lambda>0.27$, 0.09 and 0.15~fm$^{-3}$, respectively. 
These
substantially different numbers reveal the large uncertainties at high
density in the predictions of the considered functionals. 
These three functionals  
have been used before in Skyrme-Hartree-Fock 
calculations of single $\Lambda$-hypernuclei from C to Pb, and the resulting 
$\Lambda$-single particle levels have been confronted in detail to 
experimental data~\cite{vid01}. The single particle levels of the heaviest $\Lambda$-hypernuclei,
which should in principle be reasonably well described in a mean-field calculation,
were seen to be systematically over-bound by DF-NSC97a, and slightly under-bound 
by DF-NSC89, while DF-NSC97f nicely reproduces the data. 
In this sense, we can consider that the $\Lambda$ potentials in uniform
matter $v_{\Lambda}^{(N), unif}$ presented in Fig.1 span the range compatible
with $\Lambda$-hypernuclear data, even if the functionals appear to be
probably too attractive at high density \cite{vid01}.

These functionals have also been extended to other hyperonic channels,
namely $\Sigma$ and $\Xi$, assuming $SU(3)$ symmetry, and exploited in
BHF matter calculations~\cite{Stoks2000}.  However, these
extrapolations of the $N-\Lambda$ channel to the rest of the octet, 
have been severely questioned in recent years due to the difficulty of
standard functionals to predict the existence of very massive neutron
stars~\cite{Weissenborn2012,Oertel2015}. For this reason we will not
consider these extensions here.  It is however interesting to note
that all functionals predict that, for a sufficiently high number of
$\Lambda$, the $\Lambda + \Lambda \rightarrow \Xi+N$ decay channel
should open and $\Xi$ should contribute to the multistrange nuclei.
The threshold $\Lambda$ fraction for $\Xi$ appearance is estimated to
be $\rho_\Lambda/\rho_N=0.17$ for DF-NSC89 the only functional that
predicts a $\Xi$ potential in symmetric matter $V_\Xi=-15$
MeV~\cite{vid01}, compatible with experimental
data~\cite{Khaustov2000}. It should be noted that this criterion is
largely influenced by the value of the $\Xi$ potential in symmetric
matter, as expected, since for instance the threshold $\Lambda$
fraction for $\Xi$ appearance is estimated to be 
$\rho_\Lambda/\rho_N=0.15$ and 0.08~fm$^{-3}$ for the DF-NSC97a and
DF-NSC97f functionals having more attractive $\Xi$
potentials~\cite{vid01}.

This means that the results we will get for highly strange $\Lambda$-hypernuclei have to be taken with caution. 
In particular, driplines associated to a lambda fraction $\rho_\Lambda/\rho_N=0.17$ 
have to be considered as a lower bound, because even more strangeness would be compatible
with bound systems if  $\Xi$ would be accounted for.

\begin{figure}[tb]
\begin{center}
\includegraphics[width=0.5\textwidth]{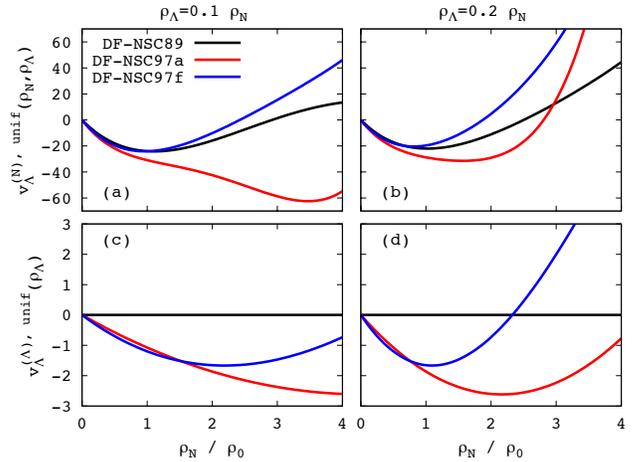}
\caption{(Color online) Comparison of $v_\Lambda^{(N), unif}$ (top panels: a, b) and $v_\Lambda^{(\Lambda), unif}$ (bottom panels: c, d),
as a function of the nucleon density $\rho_N$ (in units of the saturation density $\rho_0$) for the 
functionals DF-NSC89, DF-NSC97a and DF-NSC97f and a  $\Lambda$ fraction  set to 10\% (left panels: a, c) and 20\%
(right panels: b, d).}
\label{fig:vl2}
\end{center}
\end{figure}

The two terms $v_\Lambda^{(N), unif}$ and $v_\Lambda^{(\Lambda),
unif}$ contributing to the potential 
$v_\Lambda^{unif}$~(Eq. (\ref{eq:vlunif})) are compared in
Fig.~\ref{fig:vl2}, as a function of the nucleonic density $\rho_N$
and for different proportions  of $\Lambda$. As expected, for a
sufficiently large amount of $\Lambda$, the term $v_\Lambda^{(N),
unif}$ becomes repulsive at high nucleonic density. At low density the functionals DF-NSC97a and
DF-NSC97f predict very close potentials $v_\Lambda^{(\Lambda), unif}$,
but as the nucleonic density increases the predictions of
these two functionals increasingly deviate.
Since the dominant term in $\alpha_9$ is
attractive, the potential $v_\Lambda^{(\Lambda), unif}$ will finally
curve down at very high densities, but this occurs in physical situations
which will be hardly met in nature. Indeed, for the functional DF-NSC97a, the potential 
continuously increases up to $\rho_N\approx 10\rho_0$ and for a fraction of hyperons
$\leq 50$\%. For DF-NSC97f, the effect of $\alpha_9$ is already curving
down the potential for $\rho_N\approx 8 \rho_0$ and 20\% of $\Lambda$.
In pure $\Lambda$ matter, the potential $v_\Lambda^{(\Lambda), unif}$
becomes repulsive between $\rho_0/2$ and $\rho_0$. The functional
DF-NSC97f predicts also a 
decreasing potential starting from
$1.5\rho_0$ while this behavior is pushed to much larger values (beyond 20$\rho_0$)
with DF-NSC97a.

These differences between the functionals clearly show that the high
density properties of $\Lambda$-hyper-matter are largely unconstrained.  From
the phenomenological point of view, the existence of very massive
neutron stars (about 2M$_\odot$~\cite{Demorest2010a}) requires a
strongly repulsive $v_\Lambda^{unif}$ potential at high density, which
corresponds to a low value of the $\alpha$ parameter, in terms of the
elementary Nijmegen interactions. In principle, the very existence of
such massive stars could be used as an extra constraint for the
functional, and was often considered in the relativistic
mean-field literature (see for instance the recent
work~\cite{Sedrakian2014} and Refs. therein). However, due to the presence of different
hyperon species in the core of massive neutron stars, with highly
unknown interaction couplings, it is not straightforward to convert this
qualitative statement into a sharp constraint on the 
$v_\Lambda^{unif}$ potential. More specifically, it should be noted
that no BHF calculation is presently able to reproduce the empirical
observation of two solar mass neutrons stars. This contradiction
between our knowledge of supra-nuclear matter and observation is
usually called the hyperonization puzzle~\cite{Sedrakian}.

Finally, it can be noted from Fig.~\ref{fig:vl2} that the 
$v_\Lambda^{(\Lambda), unif}$ term is much smaller than
$v_\Lambda^{(N), unif}$: $v_\Lambda^{(\Lambda), unif}$ contributes to
less than 10\% to the total potential $v_\Lambda^{unif}$. In the
functionals DF-NSC97a and DF-NSC97f, the contribution of the term
$v_\Lambda^{(\Lambda), unif}$, induced by the $\Lambda\Lambda$
interaction, to the properties of hyperonic matter is therefore
expected to be rather weak. As discussed above, this result might not
be entirely physical, since the $\Lambda$-$\Lambda$ interaction in the
NSC97 models is not fitted on experimental data, but only extrapolated
from the $N-\Lambda$ interaction. The $\Lambda\Lambda$ term of the
functional can thus be phenomenologically adjusted as explained in
the next section.

\subsection{Phenomenological correction to the term $\epsilon_{\Lambda\Lambda}$}
\label{sec:modLL}

The term $\epsilon_{\Lambda\Lambda}$ defined by Eq.~(\ref{e_LL}) is
generated by the $\Lambda\Lambda$ interaction: only the functionals DF-NSC97a and
DF-NSC97f for which the $\Lambda\Lambda$ interaction has been included
in the BHF calculation, have a non-zero $\epsilon_{\Lambda\Lambda}$
\cite{cug00,vid01}. 

However, large uncertainties remain on the $\Lambda\Lambda$
interaction, as discussed above. The functionals introduced in section
\ref{sec:func}, based on an SU(3) extrapolation of the
$N-\Lambda$ bare interaction to the $S=-2$
channel~\cite{Maessen1989,Stoks1999}, do not lead to a satisfactory
description of the binding energy of double-$\Lambda$
hypernuclei~\cite{vid01}, which is the only empirical information that
we have on $\Lambda-\Lambda$
couplings.~\cite{Franklin95,Aoki09,Ahn13}. Moreover, Fig.~\ref{fig:vl2}
shows that, though the DF-NSC97a and DF-NSC97f functionals in the
$\Lambda-\Lambda$ channel are very similar at low density, they
drastically differ above saturation. In view of the already mentioned
hyperonization puzzle, it is therefore reasonable to consider that an even
larger uncertainty in the high density behavior has to be associated
to the term $\epsilon_{\Lambda\Lambda}$.

In the following, we therefore propose to modify the 
$\epsilon_{\Lambda\Lambda}$ term taking into account phenomenological
arguments to deal with the uncertainty on the microscopic BHF results. 
In practice, the values of the 
$\alpha_7$,$\alpha_8$, $\alpha_9$ parameters of the original functional~(\ref{functional}) 
are replaced by different values $\tilde{\alpha}_7$, $\tilde{\alpha}_8$, 
$\tilde{\alpha}_9$ defining the new 
$\tilde{\epsilon}_{\Lambda\Lambda}$ term,
\begin{equation}
\tilde{\epsilon}_{\Lambda\Lambda}=-\left(\tilde{\alpha_7}-\tilde{\alpha_8}\rho_\Lambda+\tilde{\alpha_9}\rho_\Lambda^2\right)\rho_\Lambda^2 ,
\label{eq:tildeeLLunif}
\end{equation}
and the new potential 
$\tilde{v}_\Lambda^{(\Lambda),unif}$,
\begin{equation}
\tilde{v}_\Lambda^{(\Lambda),unif}=-2\tilde{\alpha_7}\rho_\Lambda+3\tilde{\alpha_8}\rho_\Lambda^2-4\tilde{\alpha_9}\rho_\Lambda^3 ,
\label{eq:tildeVLLunif}
\end{equation}
The determination of these parameters is based on the following phenomenological prescription:
\begin{itemize}
\item[i)] Since the $\Lambda\Lambda$ interaction is expected to be repulsive at high density (to support for instance
the observed 2M$_\odot$ neutron stars), the coefficient
$\tilde{\alpha_9}$ is taken as $\le 0$.
\item[ii)] A parameter $x$ is introduced, which represents the $\Lambda$ density 
(in units of the saturation density $\rho_0$) 
where the $\Lambda$ potential in hyperonic matter
$\tilde{v}_\Lambda^{(\Lambda), unif}$~(\ref{eq:tildeVLLunif}) changes
its sign and becomes repulsive.
\item[iii)] Finally, we impose a relation between the $\Lambda\Lambda$
interaction and the bond energy in $^6$He, obtained from the
local density approximation, that we discuss hereafter~\cite{vid01}.
\end{itemize}

Condition i) imposes $\tilde{\alpha_9}\le 0$. For convenience, we set
$\tilde{\alpha_9} = 0$, giving minimal repulsion. Fig.~\ref{fig:vl2}
shows that the contribution of the term $\alpha_9$ appears at very
high densities which are certainly never reached in neutron stars.

Condition ii) and Eq.~(\ref{eq:tildeVLLunif}) give
\begin{eqnarray}
2\tilde{\alpha}_7=3\tilde{\alpha}_8\rho_\Lambda = 3\tilde{\alpha}_8 x \rho_0,
\label{eq:eq2}
\end{eqnarray}
meaning that the parameter $x$ controls the high density behavior of $\Lambda$ matter: 
the larger $x$, the softer the EoS with hyperons.

Condition iii) is related to the bond energy $\Delta B_{\Lambda\Lambda}$ which is defined 
as~\cite{cug00,vid01},
\begin{equation}
\Delta B_{\Lambda\Lambda}(A)=-E(^{A-2}Z) + 2E(^{A-1}_{\Lambda}Z)
-E(^{A}_{\Lambda\Lambda}Z),
\label{eq:bond}
\end{equation}
where $^{A-2}Z$ is a nucleus with no hyperon,
$^{A-1}_{\Lambda}Z$ is a single-$\Lambda$ hypernucleus, and
$^A_{\Lambda\Lambda}Z$ is a double-$\Lambda$ hypernucleus.
It should be noted that, experimentally, $^6$He and related hypernuclei
provide the most accurate value of a bond energy with $\Delta
B_{\Lambda\Lambda}\simeq$ 1 MeV \cite{Aoki09,Ahn13}.  

The relation between the bond energy and the functional is derived in
Appendix~\ref{app:bond}, providing
\begin{eqnarray}
\Delta B_{\Lambda\Lambda}(A) \approx
- 2 \frac{\tilde{\epsilon}_{\Lambda\Lambda}(\rho_\Lambda(A))}{\rho_\Lambda(A)} .
\label{eq:bond32}
\end{eqnarray}
Injecting Eq.~(\ref{eq:tildeeLLunif}) into Eq.~(\ref{eq:bond32}), the following relation is deduced,
\begin{eqnarray}
2 \rho_\Lambda(A)\Big[\tilde{\alpha}_7-\tilde{\alpha}_8 \rho_\Lambda(A)\Big]=\Delta B_{\Lambda\Lambda}(A)  .
\label{eq:eq1}
\end{eqnarray}
Using Eq.~(\ref{eq:eq2}) and introducing $x_\Lambda(A)=\rho_{\Lambda}(A)/\rho_0$, the average $\Lambda$-density
in double-$\Lambda$ hypernucleus $^A_{\Lambda\Lambda}Z$,  we obtain from Eq.~(\ref{eq:eq1}),
\begin{eqnarray}
\Big[3x-2 x_\Lambda(A)\Big] x_\Lambda(A) \rho_0^2 \tilde{\alpha_8}=\Delta B_{\Lambda\Lambda}(A)  .
\label{eq:eq4}
\end{eqnarray}

We can thus express the new parameters $\tilde{\alpha}_7$ and $\tilde{\alpha}_8$
as a function of $x$, $x_\Lambda(A)$ and $\Delta B_{\Lambda\Lambda}(A)$.
 from Eqs.~(\ref{eq:eq2}) and (\ref{eq:eq4}) as,
\begin{eqnarray}
\tilde{\alpha}_7&=&\frac 32 \frac{ x \; \Delta B_{\Lambda\Lambda}(A)}{\Big[3 x -2 x_\Lambda(A)\Big] x_\Lambda(A) \rho_0} , \\
\tilde{\alpha}_8&=& \frac{ \Delta B_{\Lambda\Lambda}(A)}{\Big[3 x -2 x_\Lambda(A)\Big] x_\Lambda(A) \rho_0^2} .
\end{eqnarray}

The following expressions for the energy
density and the potential are deduced from
Eqs.~(\ref{eq:tildeeLLunif}) and (\ref{eq:tildeVLLunif}):
\begin{eqnarray}
\tilde{\epsilon}_{\Lambda\Lambda}&=&-\frac{\tilde{\alpha}_8}{2} \rho_\Lambda^2 ( 3 x \rho_0- 2 \rho_\Lambda) ,\\
\tilde{v}_\Lambda^{(\Lambda), unif}&=&-3\tilde{\alpha}_8 \rho_\Lambda ( x \rho_0 - \rho_\Lambda) .
\end{eqnarray}

Since the parameters $\tilde{\alpha}_7$ and $\tilde{\alpha}_8$ are linearly related to the bond energy, 
the energy density $\tilde{\epsilon}_{\Lambda\Lambda}$ also scales with the bond energy.
The relation between $\tilde{\alpha}_7$ and $\tilde{\alpha}_8$  and the parameters $x$ and $x_\Lambda$
is more complicated.
Let us consider the limit where $x\gg x_\Lambda$.
In this case, the parameters $\tilde{\alpha}_7$ and $\tilde{\alpha}_8$ reduce to the simpler expressions:
\begin{eqnarray}
\tilde{\alpha}_7^*=\frac{\Delta B_{\Lambda\Lambda}(A)}{2 x_\Lambda(A) \rho_0} \hbox{ and }
\tilde{\alpha}_8^*=\frac{\Delta B_{\Lambda\Lambda}(A)}{3 x x_\Lambda(A) \rho_0^2} .
\end{eqnarray}

It should be noted that $\tilde{\alpha}_7^*$ is independent of $x$
while $\tilde{\alpha}_8^*$ is inversely related to $x$.  Therefore the
parameter $\tilde{\alpha}_8$ is mostly related to the prescription
ii), modifying the high density part of the mean field, while
$\tilde{\alpha}_7$ is almost independent of $x$. 
Moreover, at the limit $x\gg x_\Lambda$,  the $\Lambda$-potential
at the density $\rho_\Lambda(A)$ is given by
$\tilde{v}_\Lambda^{(\Lambda)}(\rho_\Lambda=x_\Lambda(A) \rho_0)
\approx \tilde{v}_\Lambda^{(\Lambda)*}$, and the potential
energy-density is given by
$\tilde{\epsilon}_{\Lambda\Lambda}(\rho_\Lambda=x_\Lambda(A) \rho_0)
\approx \tilde{\epsilon}_{\Lambda\Lambda}^*$, where
\begin{eqnarray}
\tilde{v}_\Lambda^{(\Lambda)*} &=& -\Delta B_{\Lambda\Lambda}(A) \\
\tilde{\epsilon}_{\Lambda\Lambda}^* &=& -0.5 \Delta B_{\Lambda\Lambda}(A) x_\Lambda(A) \rho_0 .
\end{eqnarray}
The parameter $\tilde{v}_\Lambda^{(\Lambda)*}$ is only related to the bond energy and is independent of 
$x$ and $x_\Lambda(A)$, while the parameter $\tilde{\epsilon}_{\Lambda\Lambda}^*$ depends only of the bond energy
and $x_\Lambda(A)$, independently of the value of $x$.
At the limit where $x\gg x_\Lambda$, the potential and the energy density at $x_\Lambda(A)$ depends only of
the bond energy and $x_\Lambda(A)$, and are independent of $x$.
More generally, the following relations hold:
\begin{eqnarray}
\frac{\tilde{v}_\Lambda^{(\Lambda)}(\rho_\Lambda=x_\Lambda(A) \rho_0)}{|\tilde{v}_\Lambda^{(\Lambda)*}|}&=&-1+\frac 13\frac{x_\Lambda}{x} 
+ o\left(\left(\frac{x_\Lambda}{x}\right)^2\right) \label{eq:vll2}\\
\frac{\tilde{\epsilon}_{\Lambda\Lambda}(\rho_\Lambda=x_\Lambda(A) \rho_0)}{|\tilde{\epsilon}_{\Lambda\Lambda}^*|}&=&-1
+ o\left(\left(\frac{x_\Lambda}{x}\right)^2\right) , \label{eq:ell2}
\end{eqnarray}
where the correction goes like $x_\Lambda(A)/x$.

\begin{figure}[tb]
\begin{center}
\includegraphics[width=0.5\textwidth]{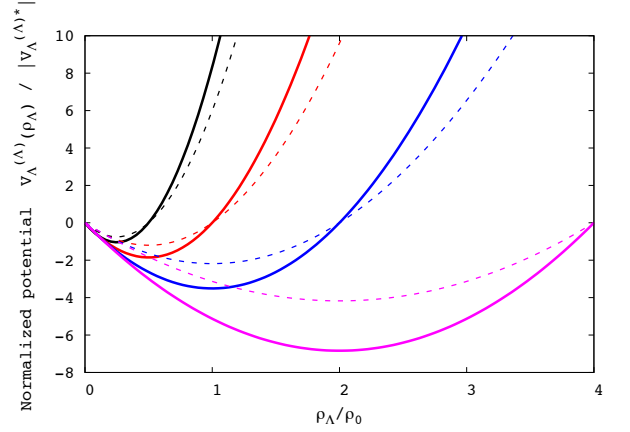}
\caption{(Color online) Normalized potential $V_{\Lambda}^{(\Lambda)}(\rho_\Lambda)/|V_{\Lambda}^{(\Lambda)*}|$
for several choices of $x$ and $x_\Lambda$, see the caption.}
\label{fig:vll1}
\end{center}
\end{figure}

\begin{figure}[tb]
\begin{center}
\includegraphics[width=0.5\textwidth]{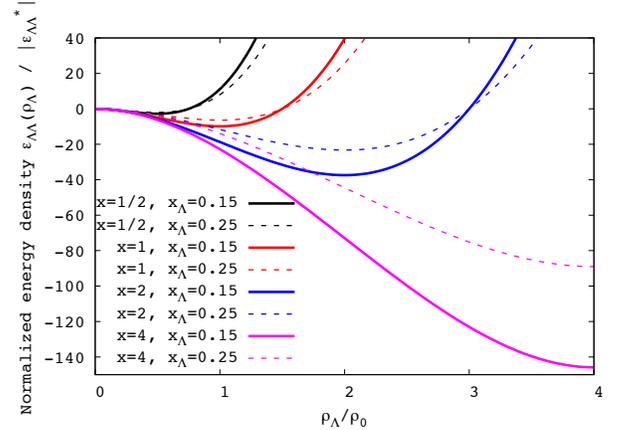}
\caption{(Color online) Normalized energy density $\epsilon_{\Lambda\Lambda}(\rho_\Lambda)/|\epsilon_{\Lambda\Lambda}^*|$
for the same parameters as in Fig.~\ref{fig:vll1}.}
\label{fig:ell1}
\end{center}
\end{figure}

Eqs.~(\ref{eq:vll2}) and (\ref{eq:ell2}) show that the scaled
$\Lambda\Lambda$ potential 
$\tilde{v}_\Lambda^{(\Lambda)}/\tilde{v}_\Lambda^{(\Lambda)*}$ and the
scaled $\Lambda$ energy
$\tilde{\epsilon}_{\Lambda\Lambda}/\tilde{\epsilon}_{\Lambda\Lambda}^*$
solely depend on $x$ and $x_\Lambda$, and are independent of the bond
energy. A complete representation of the parameter space can therefore
be obtained by representing these scaled quantities as a function of
the $\Lambda$-density, and largely varying only the two parameters
$x$ and $x_\Lambda$.  These normalized potentials (respectively energies) are
represented in Fig.~\ref{fig:vll1} (respectively \ref{fig:ell1}).

According to condition ii), the mean field potential changes its sign
for the density $\rho_\Lambda=x \rho_0$. This is well observed in
Fig.~\ref{fig:vll1}. Therefore, the effect of increasing the parameter
$x$ makes the $\Lambda$-mean field potential and energy density
softer. In a similar but weaker way, Figs.~\ref{fig:vll1} and
\ref{fig:ell1} show that the effect of the parameter $x_\Lambda$ is to
soften the mean field potential and the energy density as $x_\Lambda$
increases.

The potential and energy density behavior, displayed in
Figs.\ref{fig:vll1} and \ref{fig:ell1}, show that this choice of
parameters spans a wide range of qualitative behaviors of the
energy functional.

\subsection{Application of the energy-density functionals to $\Lambda$-hypernuclei}
\label{sec:nuclei} 

For the sake of completeness, we recall here the results obtained in
\cite{cug00,vid01}, also used in the present work, describing the
implementation of the Hartree-Fock approach in the $\Lambda$-hypernuclear case.
Nucleons and lambdas in the medium acquire an effective mass which is
generated by the momentum dependence of the interaction. The
description of finite nuclei requires to disentangle the momentum
dependent part of the in-medium potential (the correction to the
masses) from the momentum independent one, hereafter called the local
part.
The local energy density is defined by: 
\begin{eqnarray}
\frac{\hbar^2}{2m_N}\tau_N+\epsilon_{NN}(\rho_N)&=& \frac{\hbar^2}{2m^*_N(\rho_N)}\tau_N+\epsilon^{loc}_{NN}(\rho_N) ,\\
\frac{\hbar^2}{2m_\Lambda}\tau_\Lambda+\epsilon_{N\Lambda}(\rho_N,\rho_\Lambda)&=&
\frac{\hbar^2}{2m^*_\Lambda(\rho_N)}\tau_\Lambda+\epsilon^{loc}_{N\Lambda}(\rho_N,\rho_\Lambda) ,
\end{eqnarray}
which can be recasted as:
\begin{eqnarray}
&\epsilon^{loc}_{NN}& = \epsilon_{NN}-
\frac {3 \rho_N \hbar^2}{10} \left (\frac {6\pi^2 \rho_N}{g_N}\right )^{2/3} 
 \left ( \frac{1}{m^*_N}-\frac{1}{m_N}\right)
 \label{eq:elocn} \\
&\epsilon^{loc}_{N\Lambda}&=\epsilon_{N\Lambda}-
\frac {3\rho_\Lambda\hbar^2}{10} \left (\frac {6\pi^2\rho_\Lambda}{g_\Lambda}\right )^{2/3} 
\left ( \frac{1}{m^*_\Lambda}-\frac{1}{m_\Lambda}\right)
\label{eq:elocl}
\end{eqnarray}
where $\epsilon_{NN}$ is derived from the Skyrme functional \cite{Bartel2008,ben03} and
$\epsilon_{N\Lambda}$ is given by Eq.~(\ref{eq:enl}).
We have already seen that the nucleon contribution to the lambda potential, $\epsilon_{N\Lambda}$, 
is much bigger than the $\Lambda$ contribution, $\epsilon_{\Lambda\Lambda}$.
For this reason,
the contribution of the $\Lambda\Lambda$ interaction to the $\Lambda$-effective mass 
can be considered as a small corrections, and it
has been neglected in Ref.~\cite{vid01}. We can therefore write:
\begin{equation}
\epsilon^{loc}_{\Lambda\Lambda}(\rho_\Lambda)=\epsilon_{\Lambda\Lambda}(\rho_\Lambda) .
\end{equation}

\begin{figure}[tb]
\begin{center}
\includegraphics[width=0.5\textwidth]{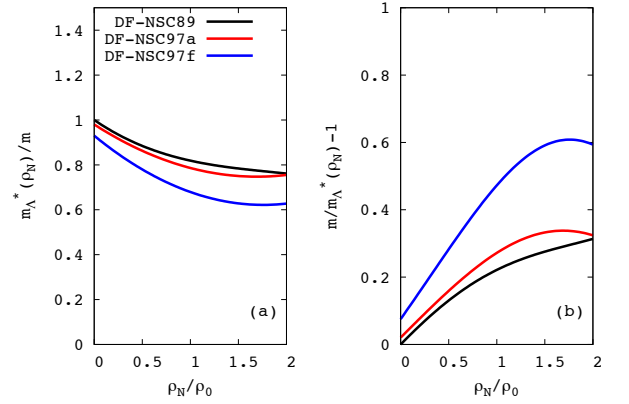}
\caption{(Color online) Ratio of the effective and bare $\Lambda$ mass 
$m_\Lambda^{*}(\rho_N)/m_\Lambda$ (a)
and $1-m_\Lambda/m_\Lambda^{*}$ (b) 
as a function of the nucleon density $\rho_N$ (in units of the saturation density $\rho_0$) 
for the functionals DF-NSC89, DF-NSC97a and DF-NSC97f.}
\label{fig:msl}
\end{center}
\end{figure}

In Eqs.~(\ref{eq:elocn}) and (\ref{eq:elocl}) the local part of the energy density requires the 
knowledge of the effective masses $m^*_N$ and $m^*_\Lambda$.
The nucleon effective mass $m^*_N$ is given from the Skyrme interaction
\cite{Bartel2008,ben03}.
The effective mass of the $\Lambda$ particles is mainly generated by the momentum
dependence of the N$\Lambda$ interaction, and it can be deduced from the BHF calculations.
The result is expressed as a polynomial in the nucleonic density
$\rho_N$ as \cite{cug00},
\begin{equation}
\frac{m_\Lambda^*(\rho_N)}{m_\Lambda} = \mu_1-\mu_2\rho_N+\mu_3\rho_N^2-\mu_4\rho_N^3 .
\label{mfit}
\end{equation}
The values for the parameters $\mu_{1-4}$ for the functional considered here
are given in Table~\ref{table:SZ2} and the density dependence of the effective mass is shown in
Fig.~\ref{fig:msl}.
Since the effective mass is only necessary in the description of finite nuclei, we have limited the
densities to values around saturation densities in Fig.~\ref{fig:msl}.
We can again observe, on the left panel, that the three functionals show a qualitatively similar behavior
for the $\Lambda$ effective mass at low density, with the stiffest model DF-NSC97f  showing 
as expected the strongest momentum dependence.

\begin{table}[t]
\caption{ Parameters of the $\Lambda$-effective mass given by Eq.~(\ref{mfit}) for the functionals considered in this paper.}
\begin{ruledtabular}
\begin{tabular}{l|ccccccccccccccccc}
Force & $\mu_1$ & $\mu_2$ & $\mu_3$ & $\mu_4$ \\
\hline
DF-NSC89~\cite{cug00,vid01}  & 1      & 1.83 & 5.33 & 6.07 \\
DF-NSC97a~\cite{vid01} & 0.98 & 1.72 & 3.18 & 0 \\
DF-NSC97f~\cite{vid01} & 0.93 & 2.19 & 3.89 & 0  \\
\end{tabular}
\end{ruledtabular}
\label{table:SZ2}
\end{table}

Given the functional form of $\epsilon^{loc}$ and $m^*$ for the nucleons and the lambdas,
Eq.~(\ref{functional}) can be rewritten in a form which explicitly
disentangle the terms coming from the local
operator $\hat\rho$ and the non-local operator $\hat\tau=\vec{\nabla}\hat\rho\vec{\nabla}$ as,
\begin{eqnarray}
\epsilon(\rho_N,\rho_\Lambda,\tau_N,\tau_\Lambda)&=&  \frac{\hbar^2}{2m^*_N(\rho_N)}\tau_N+ 
\frac{\hbar^2}{2m^*_\Lambda(\rho_N)}\tau_\Lambda \nonumber \\
&&\hspace{-1cm}+\epsilon^{loc}_{NN}(\rho_N)+\epsilon^{loc}_{N\Lambda}(\rho_N,\rho_\Lambda)
+\epsilon_{\Lambda\Lambda}(\rho_\Lambda) .
\label{functionaleff}
\end{eqnarray}

Minimizing the total energy, defined from the density functional~(\ref{functionaleff}), and using the 
Skyrme model for the nucleonic part, we obtain the usual Schr\"odinger equation ($i=N,\Lambda$),
\begin{eqnarray}
&&\Big[-\nabla\cdot \frac{\hbar^2}{2 m^*_i(r)}\nabla+V_i(r)-iW_i(r)(\nabla\times\sigma)\Big]\varphi_{i,\alpha}(r)
\nonumber \\
&&\hspace{4cm}=-e_{i,\alpha}\varphi_{i,\alpha}(r),
\end{eqnarray}
where the nucleon potential $V_N$ is defined as,
\begin{eqnarray}
V_{N}(r)&=&v_{N}^{unif}(r)+ \frac{\partial}{\partial \rho_N}\left( \frac{m_\Lambda}{m_\Lambda^*(\rho_N)} \right) \times \nonumber \\
&&\hspace{-1cm} \left( \frac{\tau_\Lambda}{2m_\Lambda}-\frac{3}{5}\frac{(3\pi^2)^{2/3}\hbar^2}{2m_\Lambda}\rho_\Lambda^{5/3}
\right) ,
\label{eq:VN}
\end{eqnarray}
The hyperon potential $V_\Lambda$ is given by,
\begin{eqnarray}
V_{\Lambda}(r)&=&v_{\Lambda}^{unif}
-\left( \frac{m_\Lambda}{m_\Lambda^*(\rho_N)} -1\right)
\frac{(3\pi^2)^{2/3}\hbar^2}{2m_\Lambda}\rho_\Lambda^{2/3} .
\label{eq:VL}
\end{eqnarray}

As in the uniform case, the hyperon potential $V_{\Lambda}$ is decomposed into
two terms,
$V_{\Lambda}=V_{\Lambda}^{(N)}+V_{\Lambda}^{(\Lambda)}$ where,
\begin{eqnarray}
V_{\Lambda}^{(N)}(r)&=& v_{\Lambda}^{(N), unif}
-\left( \frac{m_\Lambda}{m_\Lambda^*(\rho_N)} -1\right)
\frac{(3\pi^2)^{2/3}\hbar^2}{2m_\Lambda}\rho_\Lambda^{2/3} , \label{eq:VLN} \\
V_{\Lambda}^{(\Lambda)}(r)&=&v_{\Lambda}^{(\Lambda), unif} .
\label{eq:VLL}
\end{eqnarray}

The factor $(m_\Lambda/m_\Lambda^*(\rho_N)-1)$ in Eq.~(\ref{eq:VLN})
is displayed on the right panel of Fig.~\ref{fig:msl}, showing that it
is quite large around saturation density, between 0.2 and 0.5. The
modification of the lambda potential in nuclei with respect to the
uniform potential can be quite important as scales approximately with
the kinetic energy density of the lambdas.

\section{Determination of the $\Lambda\Lambda$ functional}

In the following we will use the previous formalism to perform
calculations in finite nuclei, and to
determine the parameters of the $\Lambda\Lambda$ functional.  In the case of single-$\Lambda$-hypernuclei there is no $\Lambda\Lambda$ interaction, and the corresponding terms in the functional
shall therefore
not be considered. We therefore set $\epsilon_{\Lambda\Lambda}=0$ for
single-$\Lambda$-hypernuclei, while for double and many-$\Lambda$-hypernuclei, we consider the original expression~(\ref{e_LL}) for
$\epsilon_{\Lambda\Lambda}$.

\subsection{The average $\Lambda$-density in hypernuclei}
\label{sec:ldensitynuclei}

\begin{figure}[tb]
\begin{center}
\includegraphics[width=0.5\textwidth]{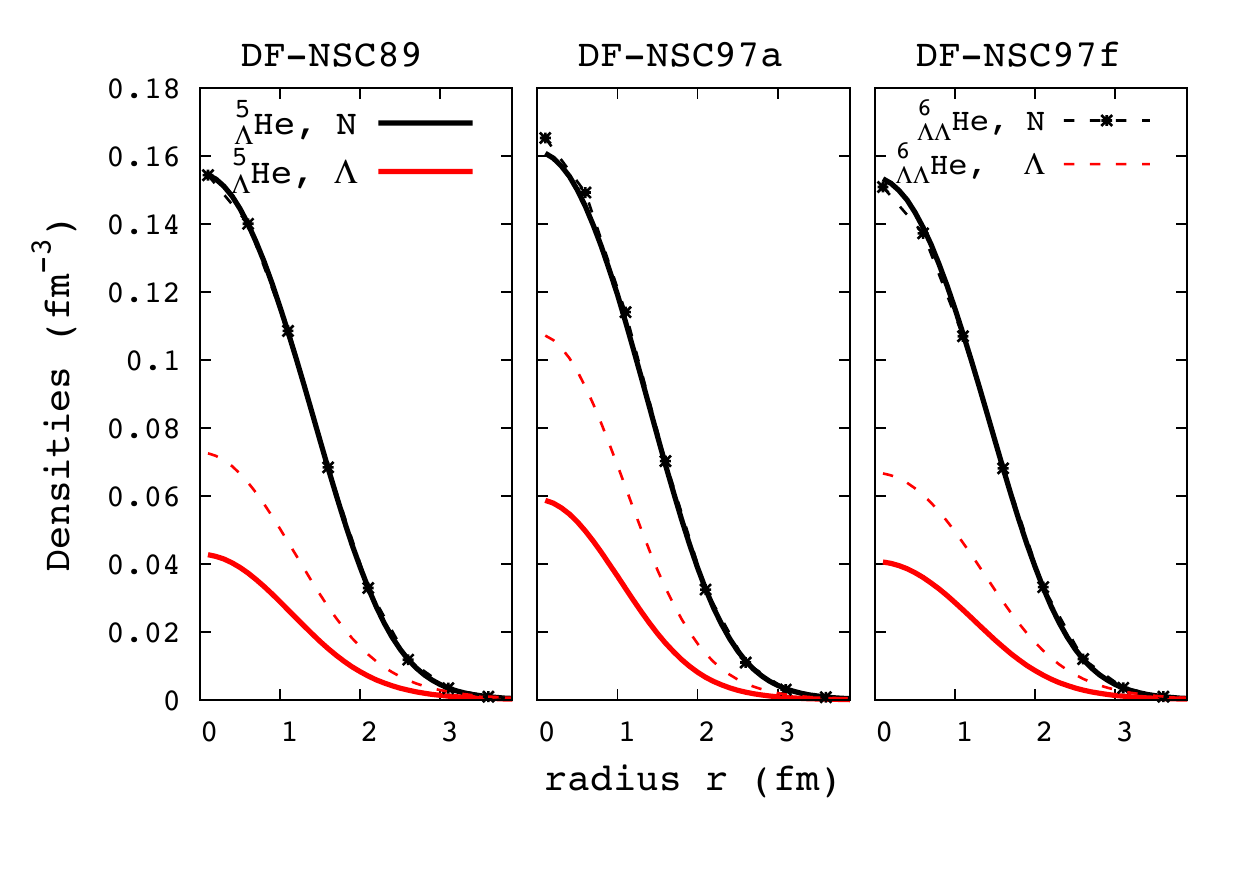}
\caption{(Color online)  Density profiles in $^4 {He}$ isotopes with the addition of one or two $\Lambda$ 
hyperons. Black (red) lines: nucleon ($\Lambda$)
density. Full (dashed) lines: one (two) $\Lambda$ is added to the $^4 He$ nucleus. 
Three different functionals, from left to right 
DF-NSC89, DF-NSC97a, and DF-NSC97f are considered.
}
\label{fig:he}
\end{center}
\end{figure}

According to Eq.~(\ref{eq:bond32}), the bond energy and the average
density in the double-$\Lambda$ hyper nucleus $^A_{\Lambda\Lambda}Z$
are closely related. The bond energy can be experimentally
constrained, but no experimental determination of the average
$\Lambda$-density exists yet. Fig.~\ref{fig:he} displays the nucleon and
$\Lambda$ densities in $^5_\Lambda$He and $^6_{\Lambda\Lambda}$He for
the functionals DF-NSC89, DF-NSC97a and DF-NSC97f, using SLy5 for the
nucleons. The average single and double-$\Lambda$-densities
(Fig.~\ref{fig:he})
show a moderate dependence on the model. 

A definition of the average $\Lambda$-density from the
$\Lambda$-density profile  is
required in order to properly define the parameter $x_\Lambda(A)$ .
Fig.~\ref{fig:he} clearly shows that
the $\Lambda$-density profile does not have a flat behavior at
the center of a  small hypernuclear system; as a consequence, the standard
deviation from the average will be quite large.
Table~\ref{table:average} displays the calculated average double-$\Lambda$ densities and standard 
deviations for the same functionals as in Fig.~\ref{fig:he}, using three different ways to estimate the average $\Lambda$-density:
\begin{eqnarray}
\langle \rho_\Lambda \rangle_\Lambda &=& \frac{\int d^3r \; \rho_\Lambda \rho_\Lambda}{\int d^3r \; \rho_\Lambda} \\
\langle \rho_\Lambda \rangle_N &=& \frac{\int d^3r \; \rho_\Lambda \rho_N}{\int d^3r \; \rho_N} \\
\langle \rho_\Lambda \rangle_T &=& \frac{\int d^3r \; \rho_\Lambda \rho_T}{\int d^3r \; \rho_T} \label{eq:rholt}
\end{eqnarray}
where $\rho_T=\rho_N+\rho_\Lambda$.
The standard deviation is defined as $\sigma_i=\sqrt{\langle \rho_\Lambda^2 \rangle_i-\langle \rho_\Lambda \rangle_i^2}$,
with $i=\Lambda$, $N$ or $T$.
The comparison between the different ways to extract the average $\Lambda$-density in Table~\ref{table:average} shows
a 15\% deviation.
As anticipated, the standard deviation is very large, almost of the order of the average value.
It is therefore difficult to properly define an average $\Lambda$-density in $^6_{\Lambda\Lambda}$He.

\begin{table}[t]
\caption{ Calculations of average $\Lambda$-densities in
$^6_{\Lambda\Lambda}$He (in fm$^{-3}$)
for the functionals DF-NSC89, DF-NSC97a and DF-NSC97f.}
\begin{ruledtabular}
\begin{tabular}{ccccccccc}
& DF-NSC89 & DF-NSC97a  & DF-NSC97f 
\\
\hline
$10^2\langle \rho_\Lambda \rangle_\Lambda$ & 2.25 & 3.69 & 2.31 
\\
$10^2\langle \rho_\Lambda \rangle_N$ & 2.64 & 3.48 & 2.65 
\\
$10^2\langle \rho_\Lambda \rangle_T$ & 2.51 & 3.55 & 2.53 
\\
\hline
$10^2\sigma_\Lambda$ & 1.97 & 2.89 & 1.85 
\\
$10^2\sigma_N$ & 1.85 & 2.74 & 1.76 
\\
$10^2\sigma_T$ & 1.90 & 2.79 & 1.80 
\\
\end{tabular}
\end{ruledtabular}
\label{table:average}
\end{table}

\begin{figure}[tb]
\begin{center}
\includegraphics[width=0.5\textwidth]{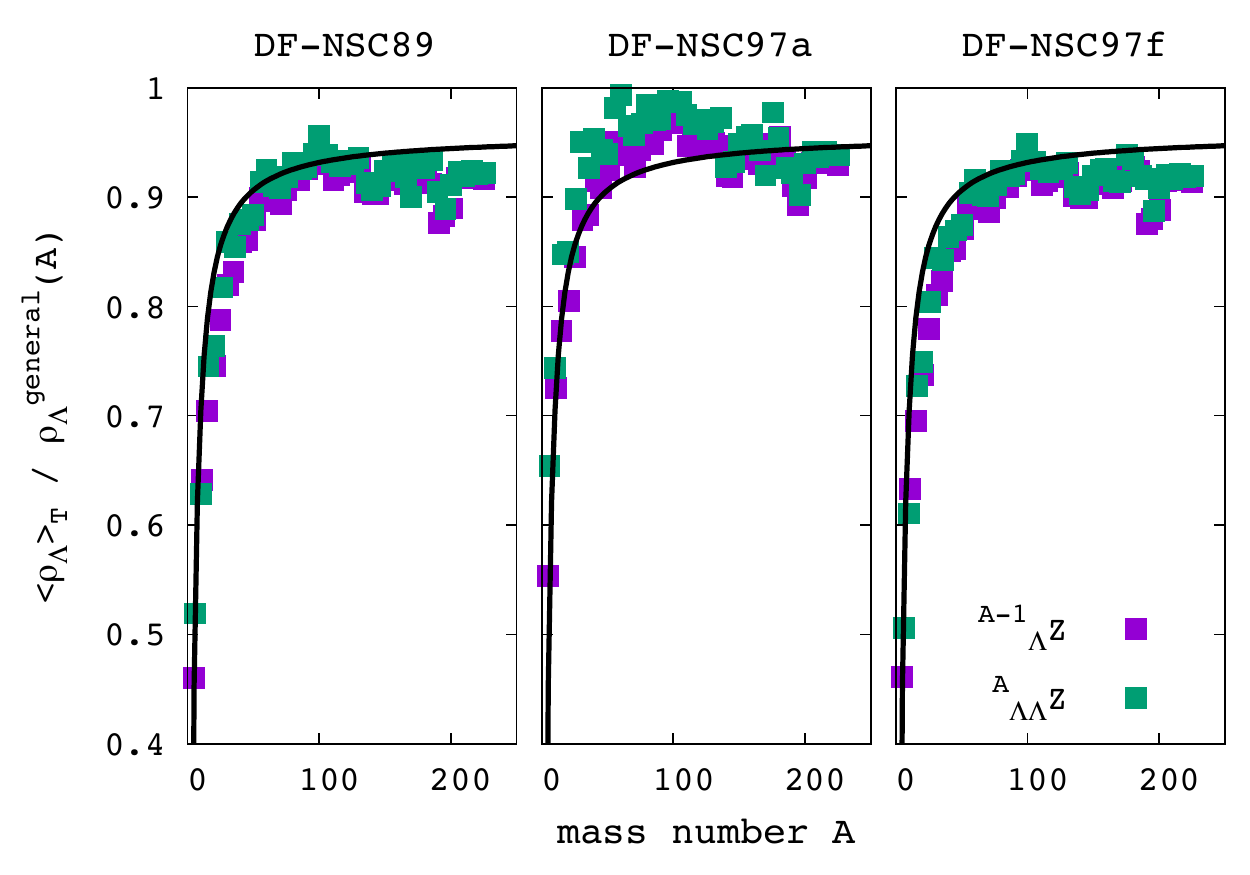}
\caption{(Color online) Ratios of the $\Lambda$ density (eq.(\ref{eq:rholt}))
over the approximate relation (eq.(\ref{genee})) as a function of baryon number 
for $\Lambda=1$ (red) and
$\Lambda=2$ (green). 
A fit is also shown.
Three different functionals, from left to right 
DF-NSC89, DF-NSC97a, and DF-NSC97f are considered.
}
\label{fig:fig4}
\end{center}
\end{figure}

A general expression for the average $\Lambda$-density in hypernuclei was
given in Ref.~\cite{vid01}, supposing that the $\Lambda$-density scales with the
nucleon density. The following general expression~\cite{vid01} was
therefore proposed:
\begin{equation}
\rho_\Lambda^{general}(A)\approx \Lambda \frac{\rho_0}{A} , 
\label{genee}
\end{equation}
where $\Lambda$=1 for single-$\Lambda$
hypernuclei and $\Lambda$=2 for double-$\Lambda$ hypernuclei.
Fig.~\ref{fig:fig4} shows the ratio of the microscopically calculated average $\Lambda$-density
$\langle \rho_\Lambda \rangle_T$, 
over the general expression (\ref{genee}). 
The general expression is rather well satisfied for large $A$, but a systematic deviation is observed
for small $A$. The fit of the deviation is also shown in Fig.~\ref{fig:fig4} and we obtain,
\begin{equation}
\langle \rho_\Lambda \rangle_T \approx \rho_\Lambda^{general}(A) \Big[  0.957-2.54 A^{-1} \Big] .
\end{equation}

This shows that Eq. (\ref{genee}) of Ref. \cite{vid01} is not valid
for A $\lesssim$ 50.  It should be noted that we have used the
expression $\langle \rho_\Lambda \rangle_T$, see Eq.~(\ref{eq:rholt}),
as a reference to estimate the average
$\Lambda$-density. Similar results are obtained with $\langle
\rho_\Lambda \rangle_N$ for large nuclei, but the definition $\langle
\rho_\Lambda \rangle_\Lambda$ gives a systematic increase by a factor
of about 2 compared to the two other definitions, in large nuclei.

\subsection{Empirical determination of the term $\tilde{\epsilon}_{\Lambda\Lambda}$}

The relation between the $\Delta B_{\Lambda\Lambda}(A)$ parameters and
$x_\Lambda(A)$ is given by Eq.~(\ref{eq:bond32}) for a given
$\Lambda$-hypernucleus $A$. However, while the bond energy can be determined
from experimental mass measurements, the average $\Lambda$-density
have never been measured. We have seen in the previous section that
the average $\Lambda$-density depends on the functional and that the
standard deviation in $^6_{\Lambda\Lambda}$He, the nucleus for which
the best measurement of the bond energy exists~\cite{Aoki09,Ahn13}, is
almost comparable with the average $\Lambda$-density. Two different
methods are used to set the value of
$x_\Lambda$:
\begin{enumerate}
\item[EmpA:] The value of $x_\Lambda(^6_{\Lambda\Lambda}$He$)$ is fixed to an average value (=1/6) in 
$^6_{\Lambda\Lambda}$He, independently of the functional.
\item[EmpB:] The value of $x_\Lambda$ is optimized to obtain a fixed bond energy ($\approx$1 MeV, consistent with 
Refs.~\cite{Aoki09,Ahn13} or $\approx$ 5 MeV, as suggested in Ref.~\cite{Franklin95})
in He for each functional.
\end{enumerate}
In prescription EmpA, the relation between the bond energy and the
functional is based on the local density approximation (see Eq.~(\ref{eq:bond32})
and App.~\ref{app:bond} for more details). It is therefore
interesting to calculate the bond energy which is obtained from
the microscopic HF calculations, in order to estimate the accuracy of the
local density approximation.
Doing so for DF-NSC89, DF-NSC97a and DF-NSC97f, and varying the
parameter $x$ from 1/2 to 4,
the difference between the bound energy set to determine the
parameters and the one 
calculated from HF calculation in $^6_{\Lambda\Lambda}$He is less than 20\%.
A larger difference is found between the parameter $x_\Lambda=1/6$ and the average $\Lambda$-density
in $^6_{\Lambda\Lambda}$He.
We conclude that the the prescription EmpA, based on the local density
approximation, cannot provide 
accurate parameters. 
This is related the analysis of the $\Lambda$-density profile in
$^6_{\Lambda\Lambda}$He which is
not flat enough to allow for the local density approximation (see
Tab.~\ref{table:average}).

\begin{figure}[tb]
\begin{center}
\includegraphics[width=0.5\textwidth]{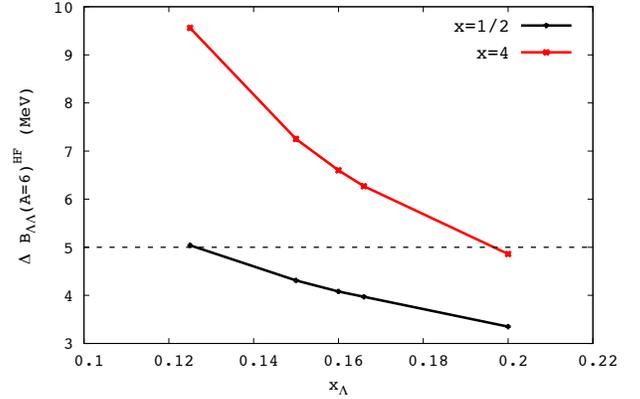}
\caption{(Color online) Relation between the parameter $x_\Lambda(^6_{\Lambda\Lambda}$He$)$
and the bond energy set to 5 MeV~(\ref{eq:bond}) for He extracted from
the HF calculation with the functional DF-NSC89 with
$x=1/2$ and DF-NSC89 with $x=4$.}
\label{fig:bond}
\end{center}
\end{figure}

In the present approach, given by the prescription EmpB, the value of $x_\Lambda$ is 
not fixed a
priori but it is varied and correlated with the bond energy
determined from the HF calculation. In such a way, the parameter
$x_\Lambda$ is treated as a variational parameter allowing to fit the
bond energy in He. 
The correlation between the parameter $x_\Lambda$ and the bond energy
$\Delta B_{\Lambda\Lambda}(A=6)^{HF}$ is shown in Fig.~\ref{fig:bond},
for the functional DF-NSC89 with $x=1/2$ and DF-NSC89 with $x=4$. The
results for the functionals DF-NSC89, DF-NSC97a and DF-NSC97f using
this empirical prescription (called EmpB hereafter) are given in
Tab.~\ref{table:presB1}, obtained from the adjustment to the bond
energy. Since a large arbitrariness is associated to the
$\Lambda$$\Lambda$ functional, the simplest polynomial form that
allows for the needed repulsion at high density, and corresponds to a
bond energy of 1 MeV in $^6$He and neighboring hypernuclei is chosen.
We will therefore use the functionals of Table~\ref{table:presB1} for
the calculations of this work. Allowing for this large variation of
the $x$ parameter, the largely unknown behavior at supersaturation is
therefore decoupled from the behavior at the very low densities
implied for the hypernuclei. 

\begin{table*}[t]
\renewcommand{\arraystretch}{1.3}
\caption{ Prescription EmpB. We present the adjustment of the parameter $x_\Lambda(4)$ to the bond energy ($\approx 1$ or 5~MeV
in $^4$He),
the values of the parameters $\tilde{\alpha}_7^{B}$ and $\tilde{\alpha}_8^{B}$, and the ratio of the $\Lambda$-density to the
saturation density in He.}
\begin{ruledtabular}
\begin{tabular}{ccccccc}
Pot. $\Lambda$N & DF-NSC89 & DF-NSC89 & DF-NSC97a & DF-NSC97a & DF-NSC97f & DF-NSC97f \\
Pot. $\Lambda\Lambda$ & EmpB1 & EmpB2 & EmpB1 & EmpB2 & EmpB1 & EmpB2 \\
$x$ & $1/2$ & $4$ & $1/2$ & $4$ & $1/2$ & $4$ \\
\hline 
& \multicolumn{6}{c}{$\Delta B_{\Lambda\Lambda}(A=6)=5$~MeV}\\
$\tilde{\alpha}_7$ & 150 & 80.82 & 150 & 70.64 & 160.48 & 100.33 \\
$\tilde{\alpha}_8$ & 1250 & 84.19 & 1250 & 73.59 & 1337 & 104.51 \\
$\Delta B_{\Lambda\Lambda}(6)^{HF}$ & 5.05 & 4.86 & 5.17 & 5.03 & 4.84 & 5.00 \\
$x_\Lambda$ & 0.125 & 0.2 & 0.125 & 0.23 & 0.115 & 0.16 \\
\hline
& \multicolumn{6}{c}{$\Delta B_{\Lambda\Lambda}(A=6)=1$~MeV}\\
$\tilde{\alpha}_7$ & 36.05 & 25.53 & 39.46 & 22.85 & 49.24 & 35.25 \\
$\tilde{\alpha}_8$ & 300.48 & 26.60 & 328.81 & 23.81 & 410.32 & 36.72 \\
$\Delta B_{\Lambda\Lambda}(6)^{HF}$ & 0.97 & 1.1 & 0.95 & 1.04 & 0.98 & 1.05 \\
$x_\Lambda$ & 0.1 & 0.125 & 0.09 & 0.14 & 0.07 & 0.09 \\
\end{tabular}
\end{ruledtabular}
\label{table:presB1}
\end{table*}

\begin{figure}[tb]
\begin{center}
\includegraphics[width=0.5\textwidth]{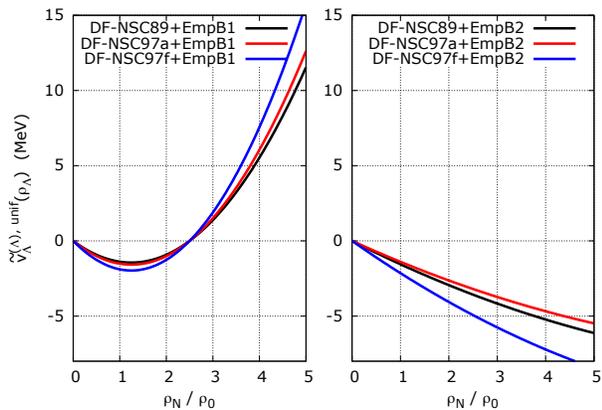}
\caption{(Color online) Comparison of the potential $v_\Lambda^{(\Lambda), unif}(\rho_N,\rho_\Lambda)$ obtained with
the prescription EmpB1 and EmpB2 (see legend) as a function of the nucleon density $\rho_N$ 
(in units of the saturation density $\rho_0$) for the functionals DF-NSC89, DF-NSC97a and DF-NSC97f.
We have fixed $\Delta B_{\Lambda\Lambda}=1$~MeV and the fraction of $\Lambda$ to be 20\%.
On the left panel, we have fixed $x=1/2$, while on the right, $x=4$.}
\label{fig:vll2}
\end{center}
\end{figure}

Finally, {Fig.~\ref{fig:vll2}} compares the potentials 
$v_\Lambda^{(\Lambda), unif}$ obtained from the functionals DF-NSC89,
DF-NSC97a and DF-NSC97f associated to the empirical prescriptions
EmpB1 and EmpB2, where the bond energy is fixed to 1~MeV. Around
saturation density, the different models shown in
{Fig.~\ref{fig:vll2}} predict negative values ranging from -1 to -2
MeV. At densities higher than normal nuclear saturation density, the
prescription EmpB1 makes the potential $v_\Lambda^{(\Lambda), unif}$
much stiffer than the prescription EmpB2. It is also interesting to
compare the empirical models shown in {Fig.~\ref{fig:vll2}} with the
initial ones displayed on {Fig.~\ref{fig:vl2}}: the empirical
prescription EmpB1 produces the stiffest potential while EmpB2 the
softest. Our empirical prescription increases the exploratory domain
of variation for the $\Lambda\Lambda$ potential, as well as includes
the initial potential.

\subsection{Calculations in $\Lambda$-hypernuclei}

In the following, we consider the three functionals DF-NSC89, DF-NSC97a and DF-NSC97f corrected
with the empirical prescriptions EmpB1 and EmpB2 associated to a bond energy of 1~MeV.
We therefore consider 6 N$\Lambda$ + $\Lambda$$\Lambda$ functionals,
corresponding to Table \ref{table:presB1}, together with SLy5
\cite{cha98} for the NN functional. 
For each N$\Lambda$ functional
(DF-NSC89, DF-NSC97a and DF-NSC97f)
the first one 
(e.g. DF-NSC89-EmpB1) 
corresponds to the lowest hyperon density in nuclear matter (x=1/2), 
whereas the second one
(e.g. DF-NSC89-EmpB2) 
corresponds to the largest hyperon density (x=4). 

It should be noted that the present work is focused on the uncertainty generated by
the $\Lambda$ related functionals on the number of bound systems. The
uncertainties generated by the NN functional itself have already been
studied on the nuclear chart and is typically of 7\% on the number of
bound nuclei \cite{erl12}. 

In the present work, the $\Lambda$-hypernuclear charts are calculated for
even-even-even $\Lambda$-hypernuclei. Since spherical symmetry is imposed, only
magic lambda number $\Lambda$-hypernuclei are considered. It should be noted
that a fully microscopic deformed approach such as Ref. \cite{zho07} is the
most accurate one to predict driplines. However such a task is numerically
very demanding, and  it is out of the scope of the present paper.

As mentioned above, the spin-orbit interaction is known to be weak in
the $\Lambda$ channel \cite{fin09}, a factor 100 lower than in the
nucleonic sector~\cite{has06}, according to experimental data. The
$\Lambda$ magic numbers are therefore expected to be close to the
harmonic oscillator ones: 2, 8, 20, 40, and 70. Calculations are
performed for $\Lambda$-hypernuclear charts corresponding to these specific
numbers of $\Lambda$.

\section{The $\Lambda$-hyperdriplines}

The $\Lambda$-hyperdripline has been studied in Ref. \cite{cug00} with a
similar HF approach, showing that the maximum number of bound
$\Lambda$ in an hypernucleus is about 1/3 of the number of nucleons.
We aim here to provide a more general study of hyperdriplines, namely
also showing the evolution of the proton and neutron driplines with
the number of hyperons. 
It is important to stress that driplines associated to a lambda fraction 
$\rho_\Lambda/\rho_N \approx 0.17$ have to be considered as a lower bound, 
because even more strangeness would be compatible
with bound systems if  $\Xi$, which are neglected in the present approach, 
would be accounted for.

\begin{figure}[tb]
\begin{center}
\scalebox{0.35}{\hspace{-4cm}\includegraphics{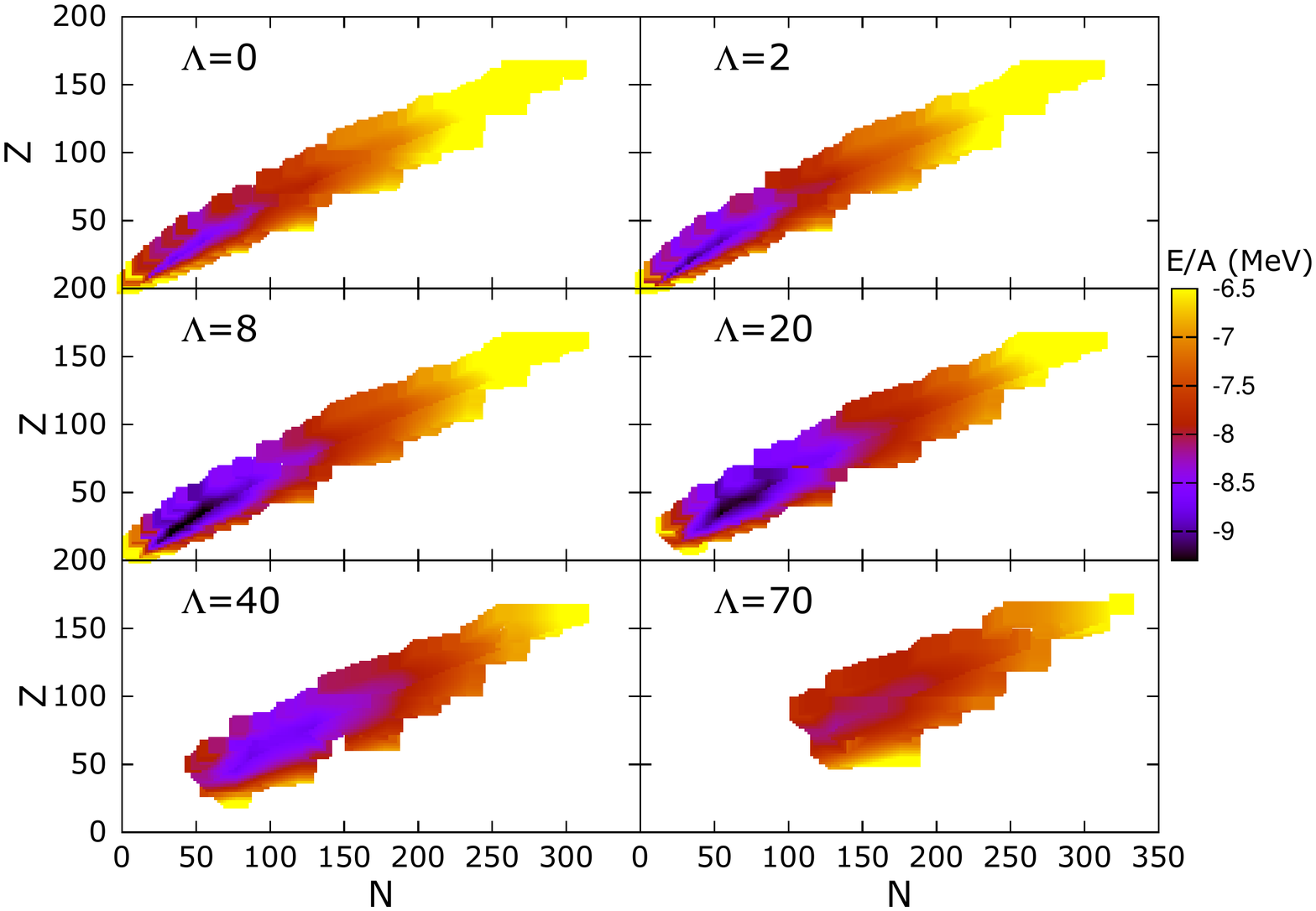}}
\caption{(Color online) $\Lambda$-Hypernuclear charts for magic $\Lambda$ numbers calculated
with the DF-NSC89+EmpB1 functional.}
\label{fig:chart}
\end{center}
\end{figure}

Figure \ref{fig:chart} displays the microscopically calculated
$\Lambda$-hypernuclear charts for $\Lambda$=0, 2, 8, 20, 40 and 70 using the 
DF-NSC89+EmpB1
functional. Adding $\Lambda$'s to a nucleus increases the binding
energy for  $\Lambda<40$, but conversely decreases it for $\Lambda$=40 and 70. This
is due to the balance between the attractive $\Lambda$N and
$\Lambda$$\Lambda$ functionals and the progressive energy filling of
the $\Lambda$ states in the mean $\Lambda$ potential. Fig.
\ref{fig:chart} also shows that with a large number of lambdas, the
corresponding hypernuclear chart is shifted towards larger N, Z values.
This effect is mainly due to the $\Lambda$N functional and is related
to the maximum fraction of 1/3 of hyperons with respect to nucleons,
as mentioned above. 
For instance the exotic $^{190}$Th core becomes bound
in the presence of 70 $\Lambda$. A similar stabilization of exotic
nuclear core has been predicted in \cite{Schaffner1994}.

\begin{figure}[tb]
\begin{center}
\scalebox{0.35}{\hspace{-1cm}\includegraphics{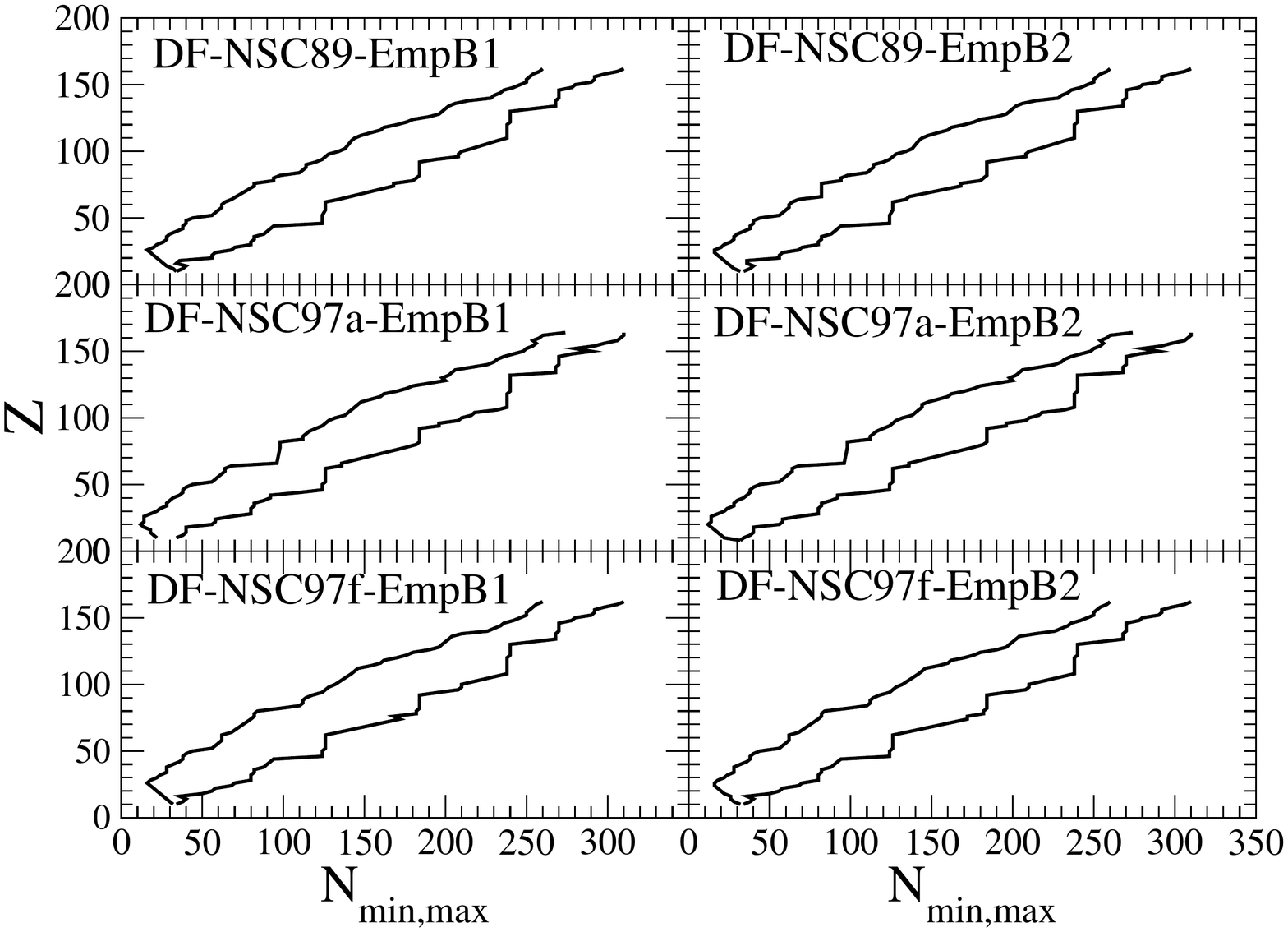}}
\caption{(Color online) Minimal (upper line) and maximal (lower line) neutron
number value for hyper nuclei with $\Lambda$=20 using the 6 functionals
described in Table \ref{table:presB1}.}
\label{fig:drip}
\end{center}
\end{figure}

To get a more accurate estimation of the dripline, Fig. \ref{fig:drip}
displays the dripline neutron numbers for each Z value in the case of
$\Lambda$=20 and for the 6 ($\Lambda$$\Lambda$+$\Lambda$N)
functionals that we consider. The dripline is here defined when the
chemical potential becomes positive.
The results are rather similar among the various
functionals. 
This shows that despite the very large differences among these functionals both in the
$N\Lambda$ and $\Lambda\Lambda$ channels, the error bars are not so large for the
drip-line determination, contrarily to what could be expected. This may
be due to the narrower density range corresponding to finite hyper-nuclei
compared to hyper-nuclear matter, since the 6 functionals provide similar potentials
for subsaturation densities.

The number of bound even-even-even $\Lambda$-hypernuclei found for $\Lambda$ $\leq$
70 and Z $\leq$ 120 are given in Table \ref{table:nee}. The dispersion due to the
uncertainty on the $\Lambda$-related functional is rather small: the average
total number of even-even-even hypernuclei for $\Lambda$=0, 2, 8, 20, 40, and 70 is
9770 $\pm$ 429. Interpolating the number of even-even-even hypernuclei
on the whole hypernuclear chart, provides 59538 $\pm$ 4020 hypernuclei
with a larger relative systematic uncertainties, which can be inferred
from the interpolation procedure. The interpolation is linearly
performed between the calculated magic $\Lambda$ hypercharts. 

We stress that this interpolation is only done with the purpose of 
estimating of the total number of bound systems. In order to have 
a detailed description of the non-magic hypernuclear chart, taking into account
deformations would be essential~\cite{zho07}.

The corresponding uncertainty is calculated from the dispersion of the
total number of even-even-even hypernuclei obtained with the six
($\Lambda$$\Lambda$+$\Lambda$N) functionals. It should also be noted
that magic nuclei are specific nuclei in the hypernuclear channel and
therefore the interpolation procedure is not optimal. Considering the
uncertainty on the bond energy (between 1 MeV and 5 MeV as explained
in the previous section), the number of estimated even-even-even
nuclei is 61460 $\pm$ 4300. The total number of hypernuclei,
considering the odd ones is therefore 491680 $\pm$ 34400. If the
uncertainty of the NN functional \cite{erl12} is included (here in a
decorrelated way), the total number of hypernuclei with $\Lambda$ $\leq$
70 and Z $\leq$ 120 is 491680 $\pm$ 59000.

\begin{table*}[t]
\renewcommand{\arraystretch}{1.3}
\caption{Number of bound even-even-even $\Lambda$-hypernuclei for $\Lambda$ $<$ 70 and Z
$<$ 120}
\begin{ruledtabular}
\begin{tabular}{l|cccccc}
 & DF-NSC89 & DF-NSC89 & DF-NSC97a & DF-NSC97a & DF-NSC97f & DF-NSC97a\\
Number of nuclei & +EmpB1 & +EmpB2 & +EmpB1 & +EmpB2 & +EmpB1 & +EmpB2\\
\hline
$\Lambda$=0 & 1578  &    1578 &   1573  &    1573  &    1578 & 1578 \\
$\Lambda$=2 &  1628 &         1640   &    1617   &     1619    &     1634  & 1621 \\
$\Lambda$=8 & 1647    &      1644     &  1692     &   1680    &    1750 & 1749 \\
$\Lambda$=20 & 1650    &      1681   &    1696    &    1724     &    1675 & 1683 \\
$\Lambda$=40 & 1713     &     1736   &    1961  &      1972     &    1722  & 1716 \\
$\Lambda$=70 & 1162     &     1237  &     1746   &     1886    &     1127 & 1152 \\
\hline
Total &  9378    &     9516  &    10285  &     10454    &    9486 & 9499 \\
Total interpolated & 56140  &      57287  &    64459   &    65587 & 56646 &   56809\\
\end{tabular}
\end{ruledtabular}
\label{table:nee}
\end{table*}

The relative uncertainty is therefore of about 4\% on the magic
 $\Lambda$ hypernuclear charts. It is determined using the mean value
 and the corresponding standard deviation considering the 6
 ($\Lambda$N+$\Lambda$$\Lambda$) functionals. It should be noted that
 the DF-NSC97a functional is more attractive than the other
 functionals, which is the main contribution to the uncertainties on
 the number of bound hypernuclei. If the bond energy requirement is
 changed from 1 MeV to 5 MeV, the variation among the 6 newly derived
 functionals is about 5 \%. All in all, it is safe to consider an
 upper limit of 7\% uncertainty due to the $\Lambda$$\Lambda$ and the
 $\Lambda$N functionals on the magic $\Lambda$ hypernuclear charts. In
 the case of interpolated values, the uncertainty is of about 7\%
 considering the 6 ($\Lambda$N+$\Lambda$$\Lambda$) functionals, plus 5
 \% from changing the bond energy requirement from 1 MeV to 5 MeV.
 These values should be compared to the relative uncertainty of about
 7\% when the NN Skyrme functional is changed \cite{erl12} on the
 nuclear chart. It shows that the uncertainty from the
 $\Lambda$-related functionals is not significantly larger than the
 one from the NN functional. This is due to the focused range of
 densities (i.e. below the saturation one) relevant for hypernuclei,
 as mentioned above.

Coming back to the sensitivity of the hyperdriplines to the
$\Lambda + \Lambda \rightarrow N+ \Xi$ decay channel, we mention that
within the generalized liquid-drop model proposed by Samanta \cite{Samanta2010} 
and straightforward to use it comes out that the n-rich frontier is almost unaffected
by considering $\Xi$s in addition to $\Lambda$s while the n-poor frontier
is shifted to lower N-values. Quite remarkable, according to Ref.  \cite{Samanta2010} 
along the n-poor dripline strangeness consists almost entirely out of $\Xi^-$ while
along the n-rich dripline strangeness is to a large extent made of $\Lambda$s. 
The broadening of the hypernuclear chart is proportional with the strangeness fraction.
The same qualitative behavior is expected to manifest also in the case of the present model.

\section{$\Lambda$-Hypernuclear structure}

\subsection{Binding energy: fusion and fission of $\Lambda$-hypernuclei}

It may be relevant to study how the presence of hyperons impacts the
most bound hypernuclei per baryon, which is known to be in the Fe-Ni
region in the case of nuclei.

\begin{figure}[tb]
\begin{center}
\scalebox{0.35}{\includegraphics{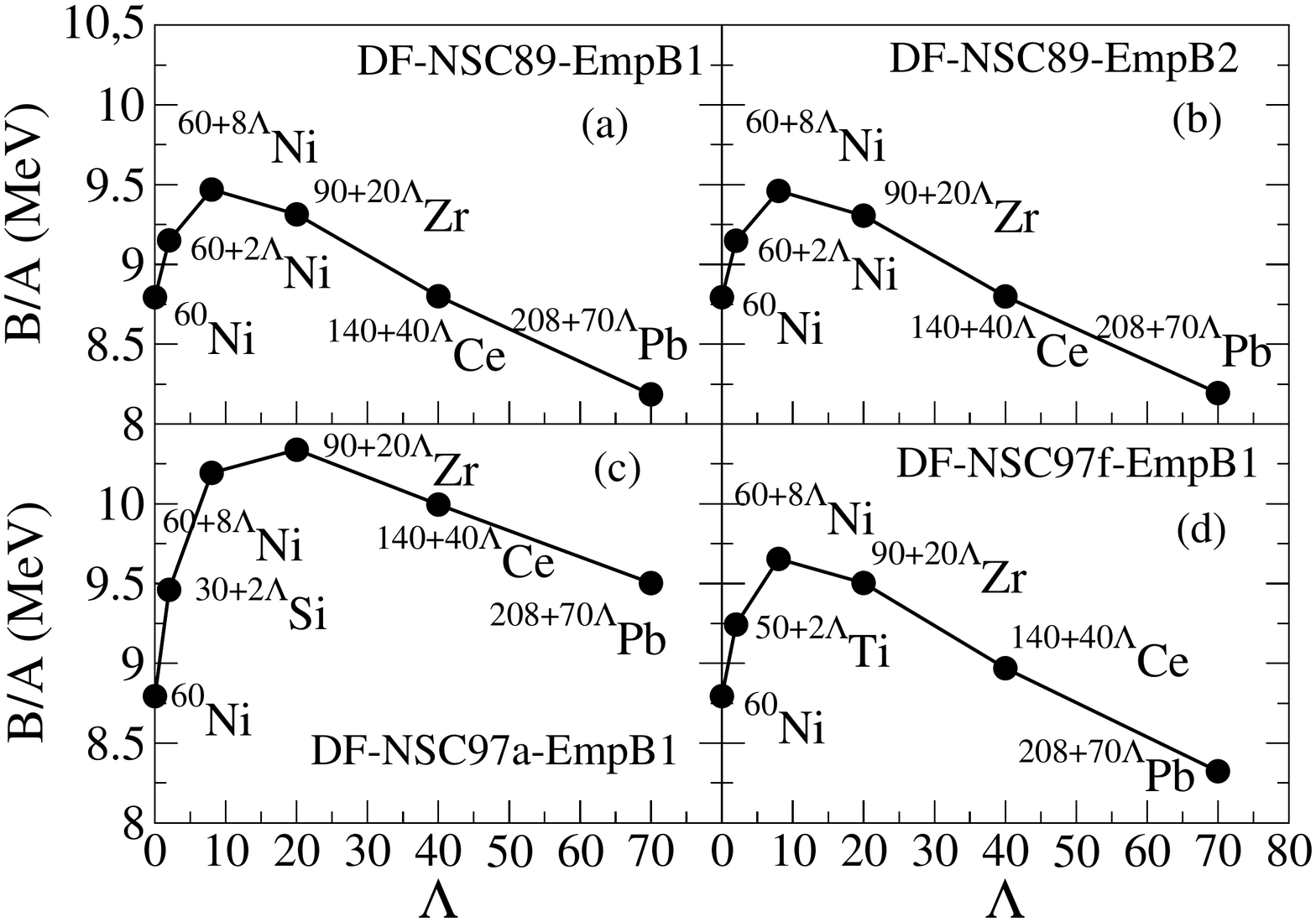}}
\caption{$\Lambda$-Hypernuclei with maximum binding energy per baryon as a
function of $\Lambda$ for the 
DF-NSC89+EmpB1 (a), DF-NSC89+EmpB2 (b), DF-NSC97a+EmpB1 (c),
DF-NSC97f+EmpB1 (d)
functionals}
\label{fig:bmax}
\end{center}
\end{figure}

Fig. \ref{fig:bmax} displays the evolution of the largest B/A value as
a function of the number of hyperons in hypernuclei calculated with 4
($\Lambda$N+$\Lambda$$\Lambda$) functionals. A remarkable agreement
between the four functionals considered is observed, which may
suggests that the theoretical uncertainty on the problem of
hypernuclear binding is relatively under control, at least in the low
density region corresponding to nuclei.  It would be very interesting
to know if this agreement is kept using relativistic functionals as in
Ref. \cite{ikr14}. In the case of DF-NSC89+EmpB1 and for $\Lambda$=2
or 8, the optimal Z value is not much changed compared to the
Nickel-Iron area. For a number of hyperons from 20 and larger the most
bound hypernucleus is obtained for larger Z, typically for Zr, Ce and
Pb. The results are qualitatively similar with the other functionals,
except in the case of DF-NSC97a+EmpB1 and DF-NSC97f+EmpB1, for
2$\Lambda$ hypernuclei: the addition of only 2$\Lambda$ makes Si and,
respectively, Ti the most bound nucleus in terms of binding energy per
baryon. This result shows the non-negligible impact of hyperons on the
binding energy of the system.

\subsection{Magicity}

Several signals of the evolution of the magic gaps along the nuclear
chart have been obtained these last decades \cite{por08}. It may
therefore be relevant to extend prediction of magicity to the
hypernuclear chart. The two protons or two neutrons gap are known to
be a relevant quantity in order to provide a first insight on magicity
features in nuclei.

\begin{figure}[tb]
\begin{center}
\scalebox{0.35}{\includegraphics{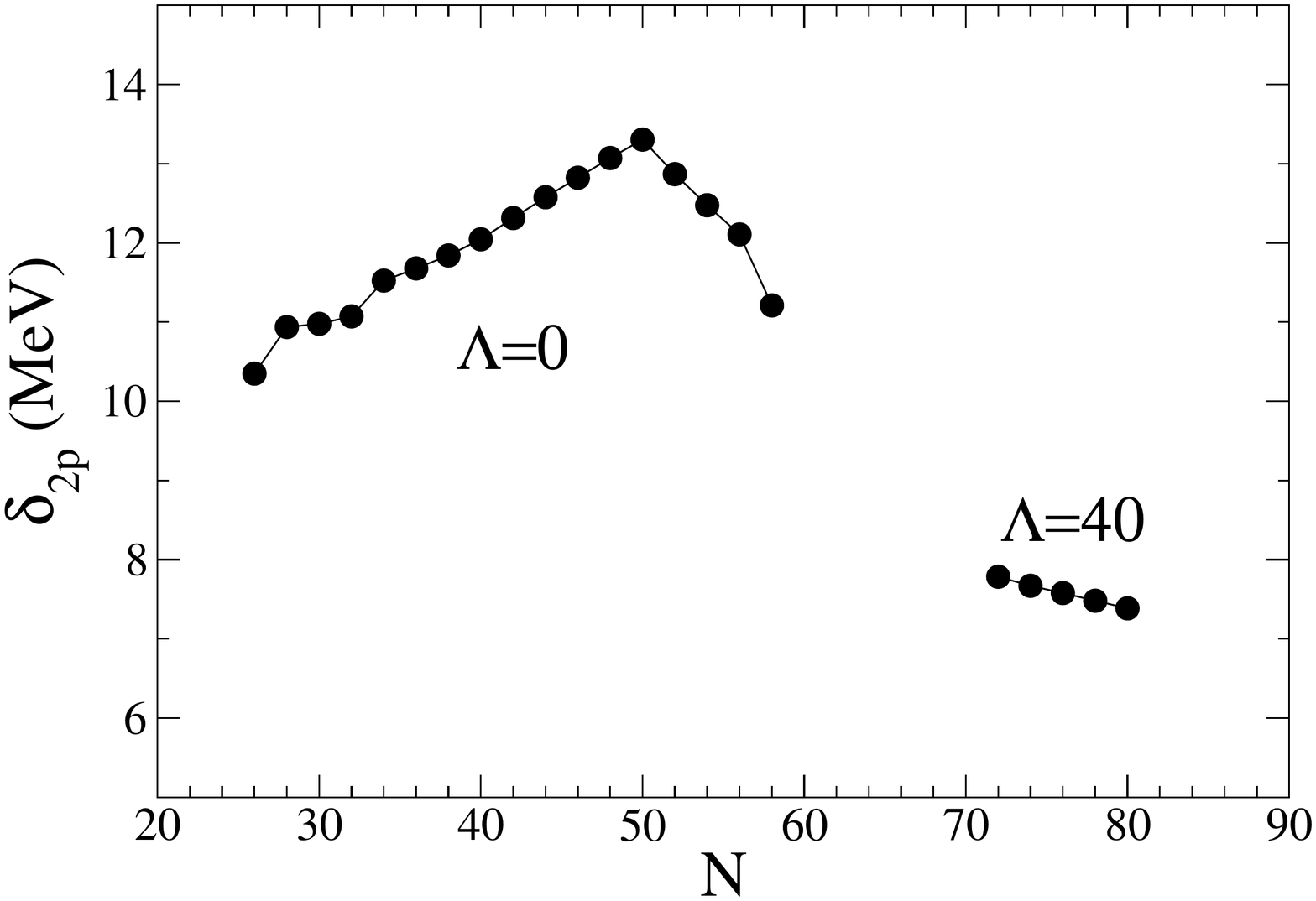}}
\caption{The 2 protons gap in the bound Nickel isotopes for
$\Lambda$=0 and  $\Lambda$=40, defined as $\delta_{2p}\equiv
S_{2p}(A,Z)-S_{2p}(A+2,Z+2)$, calculated with the DF-NSC89+EmpB1
functionnal}
\label{fig:nigap}
\end{center}
\end{figure}

It is known that shell effects can vanish in the case of very neutron
rich nuclei \cite{dob94,pen09} because of the smoothness of the
neutron skin triggering a weakening of the spin-orbit effect, and
therefore its corresponding magic numbers. 
This is the case for
hypernuclei with a low number of $\Lambda$. But in the case of
$\Lambda$=40, Figure \ref{fig:nigap} shows a restoration of the Z=28
magicity for the very neutron-rich hypernickels with 70 $<$ N $<$ 80:
the spin-orbit weakening of the Z=28 magic number for very
neutron-rich hypernuclei is reduced by the presence of hyperons. A
similar effect is observed for $\Lambda$=70 and the N=186 magic
number: it is restored for 82 $<$ Z $<$98. The present results were
obtained with 
DF-NSC89+EmpB1 and similar results are obtained with the other
functionals. A weakening of the N=28 shell closure is found for
$\Lambda$ $<$ 8 in agreement with the previous results. Also a Z=40
weakening is observed for these hypernuclei.

\begin{figure}[tb]
\begin{center}
\scalebox{0.35}{\includegraphics{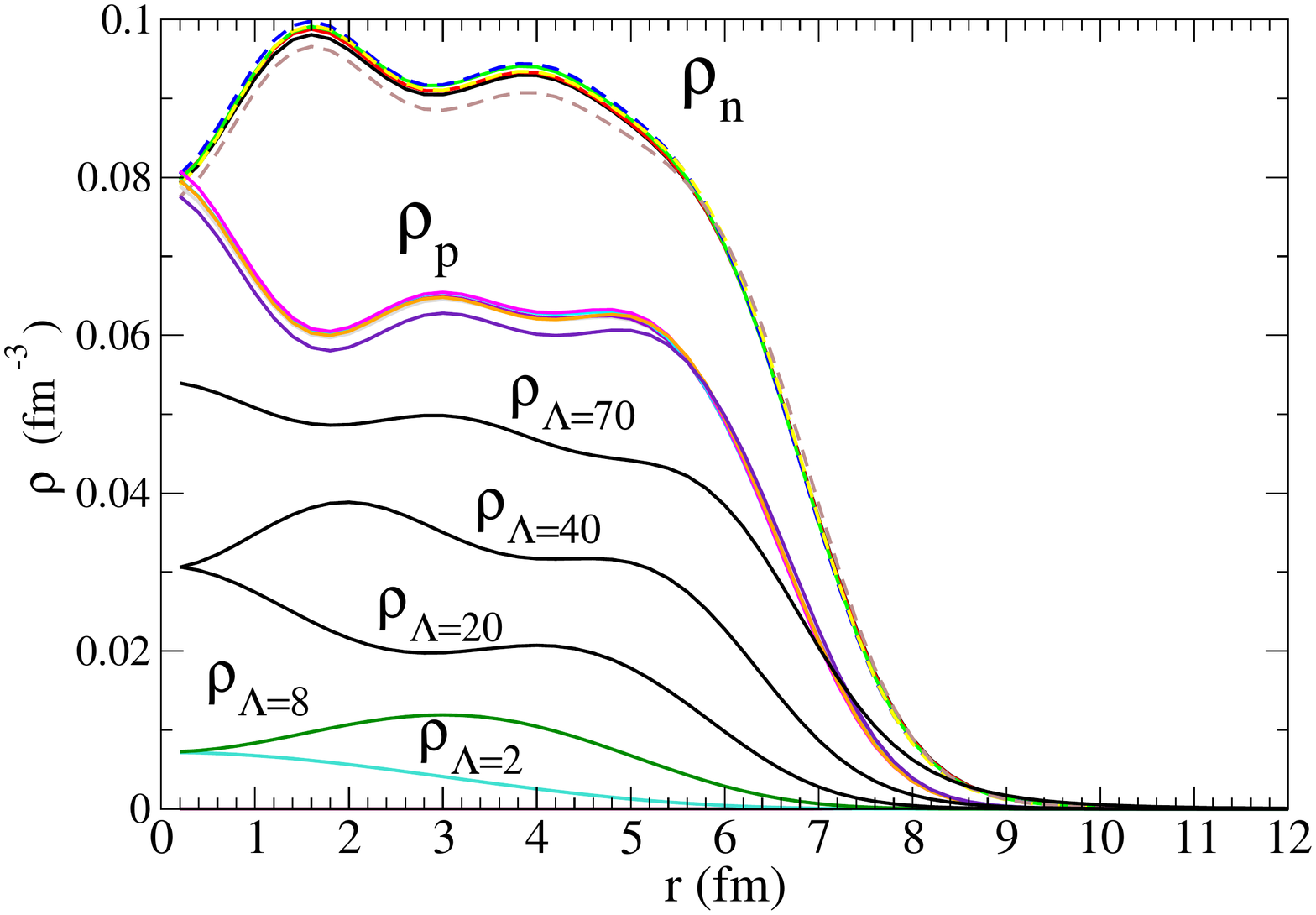}}
\caption{(Color online) Baryonic densities for the triply magic $^{208+\Lambda}$Pb
hypernucleus with $\Lambda$=2,8,20,40,70}
\label{fig:denspb}
\end{center}
\end{figure}

Fig. \ref{fig:denspb} displays the neutron, proton and hyperon
densities for the triply magic hypernucleus $^{208+\Lambda}$Pb with
$\Lambda$=2, 8, 20, 40, and 70. The proton and neutron densities are
almost not impacted by the hyperons addition, showing a relative
independence of the hyperons with respect to the nucleonic core. The
results are in agreement with Ref. \cite{cug00} where the Skyrme
parameterisation SIII was used and the $\Lambda$$\Lambda$ functional
neglected. Qualitative agreement, that is stability of radial
nucleonic mass distribution and non-monotonic evolution of strangeness
radial distributions upon increasing the number of $\Lambda$s, is also
obtained with the pioneering work of Ref. \cite{Mares1989} where
double and triple magic O and Ca hypernuclei have been addressed
within a RMF model. This validates the present approach and more
globally the microscopic prediction of hypernuclei properties. 

\subsection{Bubbles and haloes}

Bubbles and haloes effects have recently been studied in $\Lambda$-hypernuclei
\cite{ikr14}. Hyperons are more diffuse in a nucleus than nucleons.
This could be due to the weaker $\Lambda$$\Lambda$ attraction compared
to the NN one, generating an hyperon saturation density about 1/3
smaller than the nucleonic one. This may also be due to the
$\Lambda$$\Lambda$ functional which is much more intense in Ref.
\cite{ikr14} compared to the one used in the present work. It
emphasizes the importance of taking into account the bond energy in
order to constrain the $\Lambda$$\Lambda$ functional, as depicted
in section II and III.

\begin{figure}[tb]
\begin{center}
\scalebox{0.35}{\includegraphics{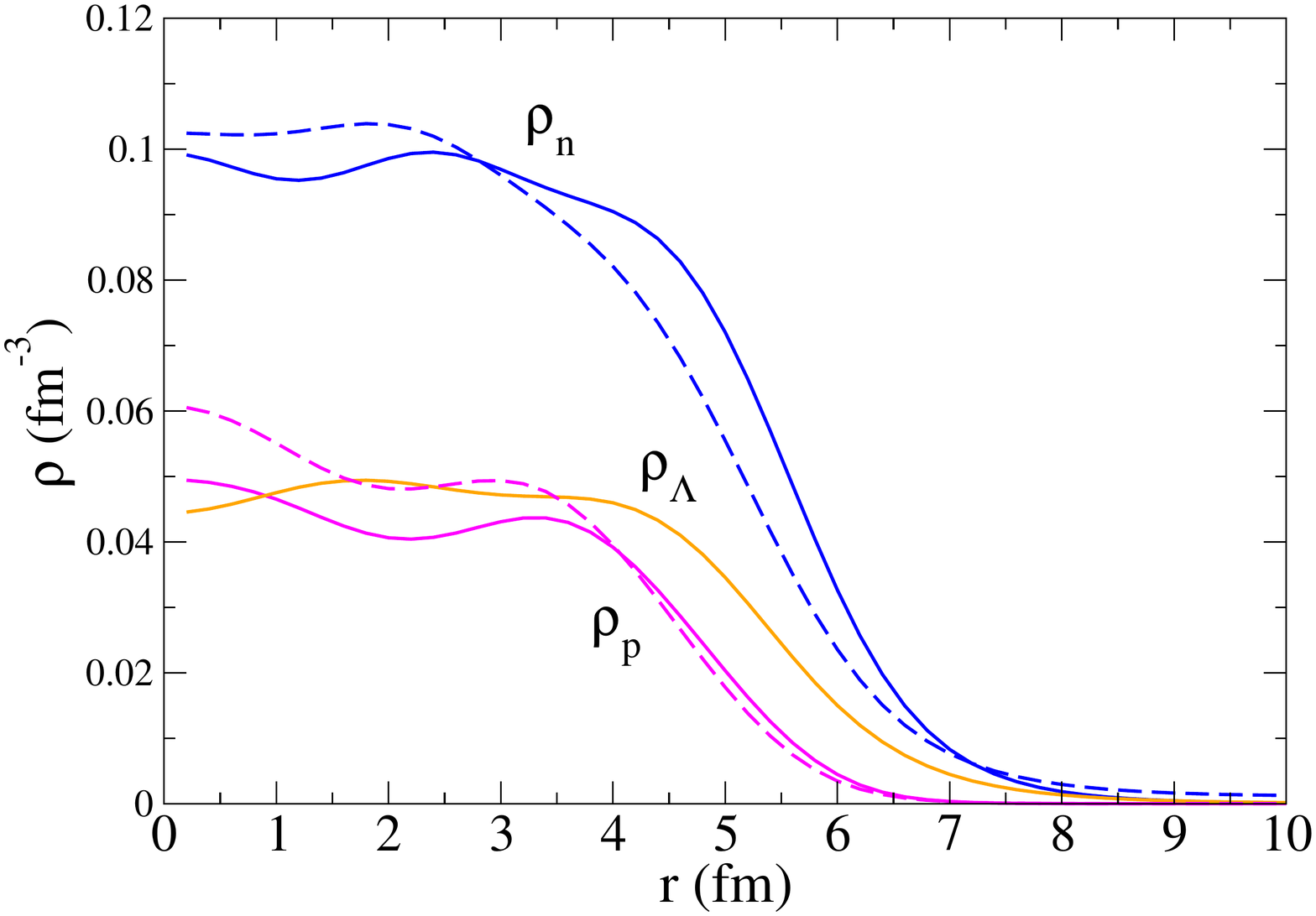}}
\caption{(Color online) Baryons densities for the magic $^{144}$Cr$_{40\Lambda}$
hypernucleus (solid lines). The neutron and proton densities for the
$^{104}$Cr nucleus are in dashed lines.}
\label{fig:denscr}
\end{center}
\end{figure}

It is well known that the addition of a single-$\Lambda$ hyperon shrinks the
nuclear core \cite{tan01}, both from predictions and from
measurements. It is therefore relevant to study the effect on the
neutron and proton densities of a large number of hyperons. We find no
large effect of the increase of the $\Lambda$ number on the proton nor
on the neutron density in $^{16}$O. In the case of
$^{104+40\Lambda}$Cr, the lambdas act like a glue between protons and
neutrons and even drive the proton to larger radii, as shown by Fig.
\ref{fig:denscr}. A similar effect is observed on the neutron density.
This is due to the fact that hyperons attract nucleons at larger
radii. It should also be noted that the halo effect in
$^{104+40\Lambda}$Cr is rather small compared to predictions using RMF
approaches \cite{ikr14}.

\begin{figure}[tb]
\begin{center}
\scalebox{0.35}{\includegraphics{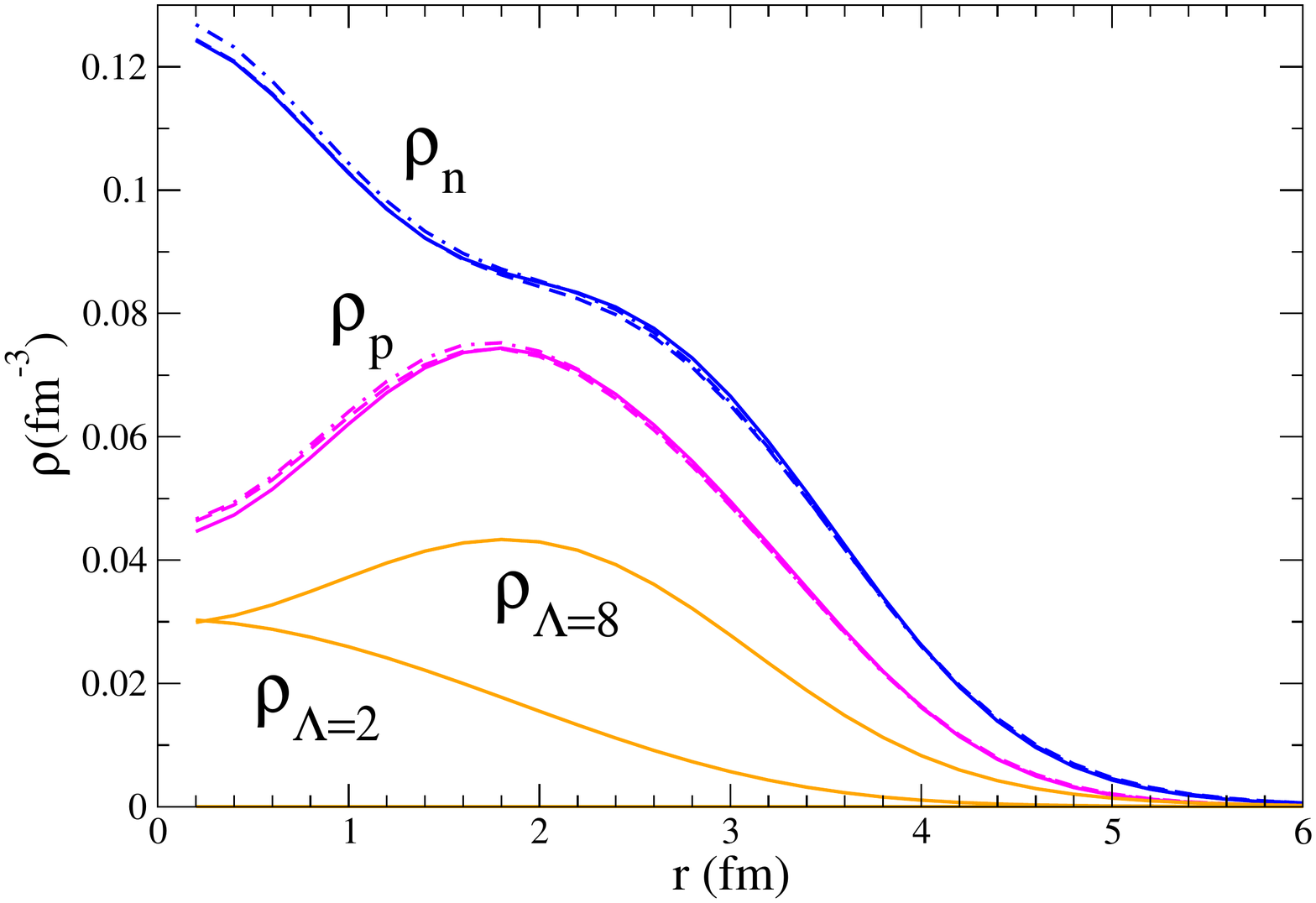}}
\caption{(Color online) Baryons densities for $^{34+\Lambda}$Si hypernucleus with
$\Lambda$=2 (dot-dashed lines for neutron and proton densities) and
$\Lambda$=8 (solid lines). The neutron and proton densities for the
$^{34}$Si nucleus are in dashed lines.}
\label{fig:denssi}
\end{center}
\end{figure}

In the case of bubbles \cite{kha08,gra09,ikr14}, there is no strong
impact of the increase of number of hyperons on the depletion. Figure
\ref{fig:denssi} displays the proton, neutrons and hyperons densities
in $^{34}$Si with no hyperons, and the addition of 2 and 8 hyperons.
As described above, the proton and neutron densities are almost not
impacted by the addition of hyperons and therefore the predicted
proton depletion in $^{34}$Si remains. This small interdependence is
at variance with relativistic calculations obtained with the RMF
approach \cite{ikr14}. 

In summary the present results show that the effect of $\Lambda$-hyperons on the
nucleon core is much weaker than in the RMF case. This may be due to
the different (N$\Lambda$+$\Lambda$$\Lambda$) functionals that are
used in the two approaches as well as to the Fock term which is not considered in the
RMF case~\cite{Massot2012}. 
It should be noted that the present
(N$\Lambda$+ $\Lambda$$\Lambda$) functionals are constrained with bond
energy requirements.

\section{Conclusions}

In this work, we have investigated the relations between $\Lambda$-hyperonic data (at low $\Lambda$-density),
and high density properties.
For that we have proceeded in two steps: first we have discussed the properties of the
functionals, based on BHF data, and proposed an empirical prescription for the $\Lambda\Lambda$ term.
Then, we have explored the hypernuclear chart allowing the functional to vary inside our domain
of uncertainty.

In the first step, we found that the low density part of the functional is well determined by the 
value of the bond energy, while the high density part is determined by the unique free parameter of the model,
which controls the density at which the $\Lambda\Lambda$ term changes its sign.  
As such we show that the bond energy controls the low
density part of the EoS solely. This means that hypernuclear data
cannot be used to determine the high density behavior of hyperonic
matter, and other constraints from neutron star physics or heavy ion
data are needed in that region. This general conclusion is certainly
independent of our model and shall be found also in other approaches.

In the second step, the $\Lambda$-hypernuclear chart for even-even-even
hypernuclei with Z $\leq$ 120 and
$\Lambda$=2, 8, 20, 40, and 70 has been calculated using the Hartree-Fock
method with Skyrme NN functional and designed N$\Lambda$
and $\Lambda$$\Lambda$ functionals. Six (N$\Lambda$
+ $\Lambda$$\Lambda$) functionals were used, optimized such as to reproduce 
the present experimental constraint of a bond
energy of 1 MeV. 
The position of the $\Lambda$-hyperdriplines is determined with a 7\%
accuracy, with comparable contribution from the uncertainty on the NN
functionals, and on the $\Lambda$ related functionals. The number of
such bound even-even-even $\Lambda$-hypernuclei is estimated to
9770 $\pm$ 429, leading to an estimation of the total number of
$\Lambda$-hypernuclei with $\Lambda <$70 and Z $<$ 120 of 491680 $\pm$ 59000.

Significant deviations from Iron-Nickel elements can be found for
$\Lambda$-hypernuclei with the largest binding energy per baryon, especially for
$\Lambda$ $\ge$ 20. The spin-orbit weakening of the neutron magicity
close to the neutron dripline is quenched the presence of hyperons.
The nucleonic core profile is not much affected by the presence of
hyperons, allowing for the persistence of the proton bubble in
$^{34}$Si with additional hyperons.

The present results shall benefit from the more and more accurate
design of the $\Lambda$-based functionals. The measurements of
$\Lambda$ and multi-$\Lambda$ hypernuclei, $\Lambda$-density profiles,
as well as $\Lambda$-$\Xi$ and $\Xi$-$\Xi$ interactions,
would greatly help to provide such critical information.

\begin{acknowledgments}
The authors thank I. Vida\~na for fruitful discussions. This work has
been partially funded by the SN2NS project ANR-10-BLAN-0503 and it has
been supported by New-Compstar, COST Action MP1304 and the Institut
Universitaire de France.
\end{acknowledgments}

\appendix

\section{Relation between the bond energy and the functional}
\label{app:bond}

For a nucleus hereafter called $(A-1)$, including one $\Lambda$ and $A-2$ nucleons, its total energy can be approximatively given
by:
\begin{equation}
E(^{A-1}_{\Lambda}Z) \approx E(^{A-2}Z) + e_\Lambda(A-1),
\label{eq:hypAm1}
\end{equation}
where $e_\Lambda(A-1)$ is the single particle energy of the $\Lambda$ state in the $^{A-1}_{\Lambda}Z$
hypernucleus, taking into account only the $N\Lambda$ interaction.
For the double-$\Lambda$ $^{A}_{\Lambda\Lambda}Z$ hypernucleus, we have also
\begin{equation}
E(^{A}_{\Lambda\Lambda}Z) \approx E(^{A-2}Z) + 2 e_\Lambda(A)+U_{\Lambda}^{(\Lambda)}(\rho_\Lambda({A}) ),
\label{eq:hypA}
\end{equation}
where $U_{\Lambda}^{(\Lambda)}(\rho_\Lambda({A}))$ is the $\Lambda$ potential term induced by the 
presence of $\Lambda$ particles.
In the present functional approach, we remind that the momentum dependence of the $\Lambda\Lambda$
interaction is neglected, and that $U_{\Lambda}^{(\Lambda)}$ is momentum independent.
Using the local density approximation, it is a function of the average $\Lambda$-density in the 
$^{A}_{\Lambda\Lambda}Z$ hypernucleus, $\rho_\Lambda({A})$.
From the BHF framework, it can be shown that~\cite{vid01}
\begin{equation}
\epsilon_{\Lambda\Lambda}=\frac 12 \rho_\Lambda U_{\Lambda}^{(\Lambda)}.
\label{eq:ellbhf}
\end{equation}

Injecting Eqs.~(\ref{eq:hypAm1}) and (\ref{eq:hypA}) into (\ref{eq:bond}), the bond energy reads,
\begin{eqnarray}
\Delta B_{\Lambda\Lambda}(A)&\approx& 2e_\Lambda(A-1) - 2e_\Lambda(A)
-U_{\Lambda}^{(\Lambda)}(\rho_\Lambda({A}) )
\label{eq:bond2}
\end{eqnarray}

It should be noted that, in order to obtain Eq.~(\ref{eq:bond2}), we have assumed the independent 
particle approximation.
Since the momentum dependence of $U_{\Lambda}^{(\Lambda)}$ is neglected,
the $\Lambda$-effective mass is the same in double-$\Lambda$ and single-$\Lambda$
hypernuclei, and depend only on $\rho_N$ as in Eq.~(\ref{mfit}).
The difference $e_\Lambda(A-1) - e_\Lambda(A)$, 
which is mostly induced by the rearrangement term in the mean field,
is calculated to be small.
We therefore approximate
$e_\Lambda(A-1)\approx e_\Lambda(A)$ to obtain
\begin{eqnarray}
\Delta B_{\Lambda\Lambda}(A) \approx
-U_{\Lambda}^{(\Lambda)}(\rho_\Lambda({A}) ) =
- 2 \frac{\epsilon_{\Lambda\Lambda}(\rho_\Lambda(A))}{\rho_\Lambda(A)} ,
\label{eq:bond3}
\end{eqnarray}
where Eq.~(\ref{eq:ellbhf}) has been used.

\section{Strangeness Analog Resonances}

Strangeness analog resonances (SAR) are similar to isobaric analog
states, with the transformation of a nucleon into an hyperon instead
of a transformation of e.g. a neutron into a proton. In Ref.
\cite{ker71}, Kerman and Lipkin studied the SAR states between nuclei
and excited states of single hyperon hypernuclei. It is interesting to
generalise this approach to multi-$\Lambda$ systems.

Kerman and Lipkin assumed that the energy difference (the
degeneracy raising) between a nucleus and the corresponding
hypernucleus state where a neutron is replaced by an hyperon, is due to i)
the mass difference between the $\Lambda$ and the neutron and ii) the
difference between the nucleonic and the hyperonic potentials
\cite{ker71}: 

\begin{equation}
\Delta E(SAR)\equiv m(^AX^*_{\Lambda})-m(^AX)=(V_{0\Lambda}-V_{0n})+(m_\Lambda-m_n)
\end{equation}

where V$_{0\Lambda}$ and V$_{0n}$ are the depth of
the hyperon and nucleons potentials, respectively, and $\Lambda$=1 here.

This corresponds to an excitation energy in the hypernucleus of

\begin{equation}
E^*\equiv m(^AX^*_{\Lambda})-m(^AX_{\Lambda})= -S_n(^AX)+(V_{0\Lambda}-V_{0n})+B_\Lambda
\end{equation}

where B$_\Lambda$ is the binding energy of the hyperon in the single
$\Lambda$ hypernucleus and S$_n(^AX)$ the one neutron separation energy
of the $^A$X nucleus.

The questions arises whether the above relation can be generalised to
multi-$\Lambda$ nuclei. A straightforward derivation allows to derive
the mass (namely the excited state) of the hypernucleus having $\Lambda$
hyperons from the mass of the initial nucleus:

\begin{equation}
\Delta E(SAR)=\Lambda\left[(V_{0\Lambda}-V_{0n})+(m_\Lambda-m_n)\right]
\label{eqsar}
\end{equation}

where $\Lambda\ge$ 1. It should be noted that the $\Lambda\Lambda$
interaction is neglected here, as well as rearrangement terms. This
could result in a significant different form of Eq. (\ref{eqsar}).
However, as shown below, the energy position of SAR states will mainly
be impacted by the nucleon vs. $\Lambda$ mass difference.

Eq. (\ref{eqsar}) shows that the SAR states in multihyperons are
expected to display a rather harmonic spectrum, since the last term in
brackets of the right-hand-side remains rather constant: there is
about 30 MeV difference between the neutron and the hyperon mean
potentials \cite{cug00}, and 170 MeV difference between the neutron
and the hyperon masses:

\begin{equation}
m(^AX^*_{\Lambda}) \simeq m(^AX) + \Lambda.200 MeV
\label{eqsar2}
\end{equation}

It should be noted that this 200 MeV constant value originates from
the saturation properties in hypernuclei, which in turn leads to a
constant difference between the nucleon and the hyperon potentials.
Moreover, in the corresponding hypernuclei, these states would
correspond to excited states located around $\Lambda$.30 MeV, which
becomes unbound already for small value of $\Lambda$. In summary, the
SAR states have been generalised to multi-hyperons hypernuclei. They
shall not correspond to bound states for most of them, but to
resonances embedded in the continuum.

\end{document}